\documentclass[a4paper]{article}
\usepackage[a4paper,top=3cm,bottom=2cm,left=2.5cm,right=2.5cm,marginparwidth=1.75cm]{geometry}
\usepackage{amsfonts}
\usepackage{bm}
\usepackage{extarrows}
\usepackage{amssymb}
\usepackage{color}
\usepackage[all]{xy}
\usepackage{graphicx}
\usepackage{braket}
\usepackage{amsmath}
\usepackage{appendix}
\usepackage{graphicx}
\usepackage[colorinlistoftodos]{todonotes}
\usepackage{amsthm}
\usepackage{mathrsfs}
\usepackage{amssymb}
\usepackage{accents}
\usepackage{extarrows}
\usepackage{makecell}
\usepackage{authblk}
\usepackage{url}
\usepackage{bbm}
\usepackage{hyperref}
\usepackage{cite}
\usepackage{eufrak}
\hypersetup{colorlinks,
            citecolor=black,
            linkcolor=black,
            urlcolor=green,
            pdftex}

\usepackage[english]{babel}
\usepackage[utf8x]{inputenc}
\usepackage[T1]{fontenc}

\newcommand{\vev}[1]{\left\langle #1 \right\rangle}

\usepackage[normalem]{ulem}

\title{Twisted geometry coherent states in all dimensional loop quantum gravity: I. Construction and Peakedness properties}
\author[1,2]{Gaoping Long \footnote{201731140005@mail.bnu.edu.cn}}
\author[1]{Xiangdong Zhang \footnote{scxdzhang@scut.edu.cn}}
\author[3]{Cong Zhang\footnote{czhang@fuw.edu.pl}\thanks{corresponding author}}
\affil[1]{Department of Physics, South China University of Technology, Guangzhou 510641, China}
\affil[2]{Department of Physics, Beijing Normal University, Beijing 100875, China}
\affil[3]{Faculty of Physics, University of Warsaw, Pasteura 5, 02-093 Warsaw, Poland}

\date{}

\begin{document}

\maketitle

\begin{abstract}
A new family of coherent states for all dimensional loop quantum gravity are proposed, which is based on the generalized twisted geometry parametrization of the phase space of $SO(D+1)$ connection theory. We prove that this family of coherent states provide an over-complete basis of the Hilbert space in which edge simplicity constraint is solved. Moreover, according to our explicit calculation,  the expectation values of holonomy and flux operators with respect to this family of coherent states coincide with the corresponding classical values given by the labels of the coherent states, up to some gauge degrees of freedom. Besides, we study the peakedness properties of this family of coherent states, including the peakedness of the wave functions of this family of coherent states in holonomy, momentum and phase space representations. It turns out that the peakedness in these various representations and the (relative) uncertainty of the expectation values of the operators are well controlled by the semi-classical parameter $t$. Therefore, this family of coherent states can serve as a candidate for the semi-classical analysis of all dimensional loop quantum gravity.

\end{abstract}

\section{Introduction}
Loop quantization method provides an approach to the quantum theory of general relativity (GR) \cite{Ashtekar2012Background, thiemann2007modern,rovelli2007quantum,RovelliBook2,Han2005FUNDAMENTAL}. As a background-independent and non-perturbative quantization scheme, this method achieves success in (1+3)-dimensional spacetime firstly. The resulting loop quantum gravity (LQG) in (1+3)-dimensional spacetime is an $SU(2)$ gauge theory including the $SU(2)$ holonomy-flux variables as conjugate pairs. This theory provided several important break-throughs. Particularly, a family of operators representing geometric observables (2-surface area, 3-region volume, inverse metric tensor) were regularized without need to subtract infinities and their spectra turned out to be discrete \cite{Ashtekar2012Background,Han2005FUNDAMENTAL}.  Moreover, by applying the loop quantization method to some symmetry reduced model, the classical cosmological and black hole singularities are resolved \cite{Ashtekar:2003hd,Ashtekar:2011ni}.
An important research interest of LQG is its semi-classical limit, which is usually proceeded based on some coherent states. The most widely used  coherent state in (1+3)-dimensional LQG is the heat-kernel coherent state, which is first proposed by Hall and further applied to general gauge field theories by Thiemann  \cite{1994The,ThiemannComplexifierCoherentStates,Thomas2001Gauge,2001Gauge,2000Gauge}. Especially, it has been shown that the heat-kernel coherent state of $SU(2)$ possesses a well-behaved peakedness property in the phase space $T^\ast SU(2)$, and the ``Ehrenfest property'' of this coherent state guarantee the coincidence between the expectation values of the elementary operators $\hat O$ and the classical evaluations of $O$ in the phase space. Based on these properties of the heat-kernel coherent state of $SU(2)$, the semi-classical limit of (1+3)-dimensional LQG is studied from a variety of perspectives \cite{Han_2020,Han_2020semiclassical,zhang2021firstorder,Zhang:2021qul}. Apart from the heat-kernel coherent state, another type of coherent state in (1+3)-dimensional LQG is introduced and applied to analyze the asymptotics of the EPRL spin foam model \cite{Rovelli_2006,Bianchi_2009,Bianchi_2010,Calcinari_2020}. Such type of coherent state is constructed by taking proper superposition over spins of the spin-networks labelled by the coherent intertwiners at vertices \cite{Calcinari_2020,Livine:2007Nsfv}. Hence, this type of coherent state is labelled by the twisted geometry parameters and called the twisted geometry coherent state \cite{Calcinari_2020}. It has been shown that the twisted geometry coherent states can be derived from the heat-kernel coherent state by selecting some special terms and then modifying them slightly . Moreover, the peakedness properties of the twisted geometry coherent state are compared with that of the heat-kernel coherent states, which shows that they have similar peakedness features.

In $(1+3)$-dimensional  LQG, the gravity and matter fields are treated separately.  However, in a well-defined quantum gravity theory, these two aspects are expected to be treated on the same footing, since the quantum-gravity effect usually appears at extremely high energy scale. Superstring theory in 10-dimensional spacetime is a candidate of such quantum gravity theory, which shows significant advantages in unifying the gravity and the other three fundamental interactions \cite{1988Superstring,1988Superstring2}. Moreover, another possibility to unify the gravity and the  other three fundamental interactions at the classical level is proposed by the classical Kaluza-Klein theory. In this theory, the unification is realized by introducing some tiny and compact extra dimensions beyond (1+3)-dimensional GR. Then the geometry of these extra dimensions will recover the matter fields by the dimensional reduction procedure \cite{Overduin:1997sri}\cite{KK2001}. The Kaluza-Klein's dimensional reduction idea has been also employed in the 10-dimensional Superstring theory to get an effective reduced theory which is expected to describe the four dimensional physics. Inspired by these quantum and classical higher dimensional gravity theories, it is worth exploring LQG in the spacetime with larger dimensions than four, where we could introduce the Kaluza-Klein's dimensional reduction idea to unify the gravity and other three fundamental interactions in a background-independent and non-perturbative quantum framework.  Pioneered by Bodendorfer, Thiemann and
Thurn \cite{Bodendorfer:Ha,Bodendorfer:La,Bodendorfer:Qu,Bodendorfer:SgI}, the loop quantum theory of GR in general $(1+D)$-dimensional spacetime (with $D\geq2$) has been developed. The quantization bases on the connection formulation of $(1+D)$-dimensional GR in the form of the $SO(D+1)$ Yang-Mills theory, where the kinematic phase space is coordinatized by the spatial $SO(D+1)$ connection field $A_{aIJ}$ and the $SO(D+1)$ Lie algebra valued densitized vector field  $\pi^{bKL}$, and equipped with the Poisson bracket $\{A_{aIJ}(x), \pi^{bKL}(y)\}=2\kappa\beta\delta_a^b\delta_{[I}^K\delta_{J]}^L\delta^{(D)}(x-y)$.  In this formulation, the dynamics are governed by a family of first-class constraints including the $SO(D+1)$ Gauss constraint, the $(1+D)$-dimensional ADM constraints and an additional constraint called the simplicity constraint. The Gaussian and ADM constraints  are similar to that in the standard (1+3)-dimensional LQG, while the simplicity constraint, taking the form $S^{ab}_{IJKL}:=\pi^{a[IJ}\pi^{|b|KL]}$, generates an extra gauge symmetry in the $SO(D+1)$ Yang-Mills phase space. It has been shown that the symplectic reduction of the kinematic phase space with respect to the Gaussian and simplicity constraints recovers the usual ADM formulation of $(1+D)$-dimensional GR.

The loop quantization of the connection formulation of $(1+D)$-dimensional GR leads to a kinematic Hilbert space spanned by the $SO(D+1)$ spin-network states. As the functions of holonomies of the connection field over edges, these spin-network states carry the quanta associating to the flux operators which represents the fluxes of $\pi^{bKL}$ over $(D-1)$-dimensional faces. The spatial geometric operators can be defined in the space which solves the Gaussian and simplicity constraints \cite{long2020operators,Long:2020agv,Zhang:2015bxa}. The Gaussian constraint is solved by the spin-network states labelled by the gauge invariant intertwiners, and the edge simplicity constraint operator, by the spin-network states whose edges are labelled by simple representations of $SO(D+1)$ \cite{Bodendorfer:Qu,Bodendorfer:2011onthe}. Besides the edge simplicity constraint, there is still the anomalous vertex simplicity constraint which is weakly solved by the spin-network states labelled by the simple coherent intertwiners \cite{long2019coherent}, referred to as the simple-coherent-interwiner spin-network states. In fact, the simple coherent intertwiners are constructed by the Perelomov type coherent states of $SO(D+1)$ minimalizing the Heisenberg uncertainty of the Lie algebra $so(D+1)$ which corresponds to that of the flux operators in $(1+D)$-dimensional LQG \cite{Long:2020euh}. Hence, the simple-coherent-intertwiner spin-network states are regarded as one kind of coherent states resembling the spatial internal geometry described by the fluxes. However, our interests are not limited to the spatial internal geometry. In order to study the semiclassical property of $(1+D)$-dimensional LQG, we need to consider the coherent states revealing the semiclassicality of the quantum spatial internal and external geometry simultaneously. To construct such kind of coherent state in $(1+D)$-dimensional LQG with the compact gauge group $SO(D+1)$, one could simply extend the Thiemann's procedure for $SU(2)$ case to the $SO(D+1)$ case to construct the heat-kernel coherent state of $SO(D+1)$.  However, because of the complicatedness of $SO(D+1)$, some specific calculations about the $SO(D+1)$ heat-kernel coherent state, like studying the peakedness and Ehrenfest properties, will be too difficult to proceed with.

Nevertheless, there is another way to construct the coherent state in $(1+D)$-dimensional LQG. As in the (1+3)-dimensional case, we can take advantage of the twisted geometry parametrization of the phase space $T^\ast SO(D+1)$ to construct the twisted geometry coherent state in $(1+D)$-dimensional LQG \cite{PhysRevD.103.086016}. Since the twisted geometry parameters exclude the degrees of freedom which are expected to be eliminated in symplectic reduction procedure by edge simplicity constraint, the twisted geometry coherent state will be much simpler without losing any physical degrees of freedom. Furthermore, the twisted geometry coherent state constructed through this approach make it possible to solve some unsettled difficulties. Let us explain it briefly. Similar to the $(1+3)$-dimensional case, the scalar constraint in $(1+D)$-dimensional LQG is expressed as the summation of an Euclidean term and a Lorentzian term. However, since the connection contains some pure gauge components which transforms under the gauge transformation induced by the simplicity constraint, the Euclidean term is gauge variant and an extra term must be added to offset the contribution of the pure gauge component. This extra term introduces huge obstacle in the regularization and quantization of  the scalar constraint. However, the twisted geometry parameter is gauge invariant with respect to the simplicity constraint. Thus, if the twisted geometry coherent state has well peakedness properties, one can expect that the matrix element of holonomy operator with respect to the twisted geometry coherent states will exclude the dependence on the pure gauge component. Taking of advantage of the twist geometry coherent state to calculate the quantum scalar constraint, one could omit the extra term to offset the contribution of the pure gauge component in the Euclidean term. Thus, the calculation could only involves the Euclidean term and Lorentzian term as that in (1+3)-dimensional $SU(2)$ LQG.

 In fact, the superposition type coherent state introduced in our previous work is the first example of the coherent state constructed based on the twisted geometry parametrization in $(1+D)$-dimensional LQG. It has been shown that \cite{Long:2021xjm}, this coherent state is identical to the heat-kernel one in large conjugate momentum limit and provides a resolution of identity of the solution space of the quantum edge simplicity constraint. Moreover, it is verified for the one-loop-graph case that the peakedness property of the superposition type coherent state behaves well in large conjugate momentum limit.  However, the superposition type coherent state has some critical defects.  The superposition type coherent state is constructed by taking summation of the coherent-intertwiner spin-network states weighted by a complex Gaussian function of some quantum numbers, where the center of this Gaussian function is determined by the twisted geometry parameters. Since these quantum numbers as the arguments of this Gaussian function take only positive values, the center of those state for which the peakedness property is well-behaved must be positive and far away from 0. However, to get the resolution of identity, we have to consider all Gaussian functions whose centers take values of the entire $\mathbb{R}$.  Thus the peakedness property is not satisfied by all twisted geometry coherent states. Moreover, the twisted geometry parameter space has two branches giving a two to one parametrization of the reduced phase space, which implies that the twisted geometry parametrization possesses a $\mathbb{Z}_2$ symmetry \cite{PhysRevD.103.086016}. Though the superposition type coherent states are labelled by twisted geometry parameters, they lost the $\mathbb{Z}_2$ symmetry of the the twisted geometry parametrization.
 Hence, an improvement on the construction of coherent state based on twisted geometry parametrization is desired.

In this paper, we will proposal a new family of coherent state constructed via the twisted geometry parametrization of the phase space of $(1+D)$-dimensional LQG, which will be referred to as the twisted geometry coherent state. This scheme of construction is a $SO(D+1)$ extension of the $SU(2)$ twisted geometry coherent state \cite{Calcinari_2020}. More explicitly, the twisted geometry coherent state contains the superpositions of two series of Perelomov type coherent states of $SO(D+1)$ given by the highest and lowest weights respectively, instead of only a single series of that given by the highest weight in the superposition type coherent state. The two series of Perelomov type coherent states ensure the realization of the $\mathbb{Z}_2$ symmetry of the twist geometry parametrization by the twisted geometry coherent state. Then, compared with the superposition type coherent states, the twisted geometry coherent states have the well-behaved peakedness property for a larger range of their labels. In addition, based on the twisted geometry coherent state, we will further calculate the expectation values of the basic operators and show the peakedness  in the configuration, momentum and phase space representations for general graphs, instead of the simplest one-loop graph considered for superposition type coherent state in our previous work \cite{Long:2021xjm}.

This paper is organized as follows. In section \ref{sec2}, we will review the basic structures of all dimensional LQG. Beginning with the connection dynamics of GR in $(1+D)$-dimensional spacetime, the holonomy-flux phase space and its twisted geometry parametrization will be introduced. Then, the quantum Hilbert space of all dimensional LQG spanned by the spin-network basis labelled by the coherent intertwiners will be pointed out. In section \ref{sec3},  the explicit expressions and some main properties of the heat-kernel coherent states and the superposition type coherent states in all dimensional LQG will be reviewed. Then in sections \ref{sec4}, we will construct the new family of twisted geometry coherent states and study their properties explicitly, including the over-completeness property, peakedness properties and the expectation values of the basic operators. Finally, after a short conclusion in section \ref{sec5}, we will end with some discussions and outlooks for the possible next steps of the future research.

\section{Kinematic structure of all dimensional loop quantum gravity}\label{sec2}
\subsection{The connection phase space of all dimensional loop quantum gravity}
The connection dynamics of  $(1+D)$-dimensional GR is based on the $SO(D+1)$ Yang-Mills theory, whose phase space is coordinatized by the $so(D+1)$ valued connection $A_{aIJ}$ field and its conjugate momentum $\pi^{bKL}$ on a spatial D-dimensional manifold $\sigma$.
The only non-vanishing Poisson brackets between the canonical pairs  $(A_{aIJ}, \pi^{bKL})$ is given by
\begin{equation}\label{Poisson1}
\{{A}_{aIJ}(x), \pi^{bKL}(y)\}=2\kappa\beta\delta_a^b\delta_{[I}^K\delta_{J]}^{L}\delta^{(D)}(x-y).
\end{equation}
Here we use the notation $a,b,...=1,2,...,D$ for the spatial tensorial indices and $I,J,...=1,2,...,D+1$ for the $so(D+1)$ Lie algebra indices in the definition representation.
In addition to the scalar constraints $C\approx0$ and vector constraints $C_a\approx0$ in the geometric dynamics of  $(1+D)$-dimensional GR, two kinds
of extra constraints appear for this Yang-Mills phase space, which are given as
\begin{equation}
\text{Gauss \ constraint:}\quad\mathcal{G}^{IJ}\equiv \partial_a\pi^{aIJ}+2{A}_{aK}^{[I}\pi^{a|K|J]}\approx0,
\end{equation}
\begin{equation}
\text{Simplicity \ constraint:}\quad S^{ab[IJKL]}=\pi^{a[IJ}\pi^{|b|KL]}\approx0.
\end{equation}
It has been shown that these four constraints obey a first class constraint algebra \cite{Bodendorfer:Ha}. As expected, the Gauss constraint generates the $SO(D+1)$ gauge transformation among the internal degrees of freedom, while the simplicity constraint places restrictions on  $\pi^{aIJ}$ to adapt its degrees of freedom to describe the spatial internal geometry. 
More explicitly, the simplicity constraint can be solved as $\pi^{aIJ}=2n^{[I}E^{|a|J]}$, where $n^{I}E^{a}_{I}=0, n^In_I=1$ and  $E^{aI}$ is a densitized D-frame related to the spatial metric by $qq^{ab}=E^{aI}E^b_I$ with $q$ being the determinant of $q_{ab}$ \cite{Bodendorfer:Ha}. Moreover, on the Gaussian and simplicity constraint surface, the densitized extrinsic curvature of the spatial manifold $\sigma$ can be given as
\begin{equation}
\tilde{{K}}_a^{\ b}\approx{ K}_{aIJ}\pi^{bIJ}\equiv \frac{1}{\beta}({A}_{aIJ}-\Gamma_{aIJ})\pi^{bIJ},
\end{equation}
 where $\Gamma_{aIJ}$ is the spin connection defined by $E^{aI}$ \cite{Bodendorfer:Ha}. One can decompose ${K}_{aIJ}
:=\frac{1}{\beta}({A}_{aIJ}-\Gamma_{aIJ})$ as ${K}_{aIJ}=2n^{[I}{K}_a^{J]}+\bar{{K}}_{a}^{IJ}$, where $\bar{{K}}_{a}^{IJ}:=\bar{\eta}^I_K\bar{\eta}^J_L{K}_{a}^{KL}$ with $\bar{\eta}_I^J:=\delta_I^J-n_In^J$ and $\bar{{K}}_{a}^{IJ}n_I=0$. Then, based on the Poisson bracket \eqref{Poisson1}, it is easy to verify that the component $\bar{{K}}_{a}^{IJ}$ transforms while $2n^{[I}{K}_a^{J]}$ and $\pi^{aIJ}$ are invariant along the Hamiltonian flow of the simplicity constraint on the Gaussian and simplicity constraint surface. A more detailed study has shown that the symplectic reduction of the $SO(D+1)$ Yang-Mills phase space with respect to Gaussian and simplicity constraint reproduces the ADM phase space of $(1+D)$-dimensional GR, with $\tilde{{K}}_a^{\ b}$ and $q_{ab}$ are Dirac observables with respect to Gaussian and simplicity
constraints \cite{Bodendorfer:Ha}. Especially, we note that $\bar{{K}}_{a}^{IJ}$ is a pure gauge component with respect to the simplicity constraint, which only contributes gauge degrees of freedom in this theory.
\subsection{The discrete phase space of all dimensional loop quantum gravity}
The discrete phase space in $(1+D)$-dimensional loop quantum gravity is constructed based on the spatially smeared variables, including the holonomies of connection over edges and the fluxes of conjugate momentum over the $(D-1)$-faces. Usually, the set of edges to define holonomies and $(D-1)$-faces to define fluxes are provided by a given graph and the cell decomposition dual to it respectively. Based on this construction and the Poisson algebra between connection and its conjugate momentum, the holonomy over one edge conjugates to the flux over the $(D-1)$-face traversed by the edge naturally. Then, the Poisson brackets between the pairs of holonomies and fluxes associated with the given graph compose the discrete version of the algebra \eqref{Poisson1}. Thus, the holonomies and fluxes associated with the given graph form a new discrete phase space. More explicitly, for a given graph $\gamma$ embedded in the spatial D-manifold $\sigma$, we have the new algebra consisting of the discrete variables $(h_e, X_e)\in \times_{e\in\gamma}(SO(D+1)\times so(D+1))_e$. Here the holonomy
$h_e$ is defined by $h_e:=\mathcal{P}\exp\int_e \mathcal{A}$, wherein $\mathcal{A}:=\frac{1}{2}\dot{e}^aA_{aIJ}\tau^{IJ}$, $\dot{e}^a$ is the tangent vector field of $e$, $\tau^{IJ}$ is a basis of $so(D+1)$ given by $(\tau^{IJ})^{\text{def.}}_{KL}=2\delta^{[I}_{K}\delta^{J]}_{L}$ in definition representation space of $SO(D+1)$, and $\mathcal{P}$ denotes the path-ordered product. The flux $F_e$ is defined by $F_e:=\int_{e^\star}(h\pi h^{-1})^an_ad{}^{D-1}\!S$, and the dimensionless flux $X_e:=\frac{1}{\beta a^{D-1}}\int_{e^\star}(h\pi h^{-1})^an_ad{}^{D-1}\!S$, where $e^\star$ represents the $(D-1)$-dimensional face traversed by $e$ in the dual lattice of $\gamma$, $e^\star$ has the normal $n_a$ and the infinitesimal coordinate area element $d{}^{D-1}\!S$, $a$ is an arbitrary but fixed constant with the dimension of length, and $h$ is the holonomy from the start point of $e$ to the point of integration along a path adapted to $\gamma$. Notice $SO(D+1)\times so(D+1)\cong T^\ast SO(D+1)$, so the discrete phase space $\times_{e\in \gamma}(SO(D+1)\times so(D+1))_e$ is a direct Cartesian product of $SO(D+1)$ cotangent bundles and it is called the discrete phase space of $(1+D)$-dimensional LQG on graph $\gamma$. Finally, the complete phase space of $(1+D)$-dimensional LQG is simply the union of the discrete phase spaces on all possible graphs.
Let us consider the discrete phase space associated with $\gamma$. In this space, the discretized version of Gaussian constraints read
 \begin{equation}
 G_v:=\sum_{b(e)=v}X_e-\sum_{t(e')=v}h_{e'}^{-1}X_{e'}h_{e'}\approx0
 \end{equation}
 and the discretized version of simplicity constraints consist of two sets, including the edge simplicity constraints $S^{IJKL}_e\approx0$ and vertex simplicity constraints $S^{IJKL}_{v,e,e'}\approx0$, which are given as
\begin{equation}
\label{simpconstr}
S_e^{IJKL}\equiv X^{[IJ}_e X^{KL]}_e\approx0, \ \forall e\in \gamma,\quad S_{v,e,e'}^{IJKL}\equiv X^{[IJ}_e X^{KL]}_{e'}\approx0,\ \forall e,e'\in \gamma, s(e)=s(e')=v.
\end{equation}
The Poisson brackets between $(h_e, X_e)$ take the forms
 \begin{eqnarray}
 &&\{h_e, h_{e'}\}=0,\quad\{h_e, X^{IJ}_{e'}\}=\delta_{e,e'}\frac{\kappa}{a^{D-1}} \frac{d}{dt}(e^{t\tau^{IJ}}h_e)|_{t=0}, \\\nonumber
 && \{X^{IJ}_e, X^{KL}_{e'}\}=\delta_{e,e'}\frac{\kappa}{a^{D-1}}(\delta^{IK}X_e^{JL}+\delta^{JL }X^{IK}_e-\delta^{IL}X_e^{JK}-\delta^{JK}X_e^{ IL}).
 \end{eqnarray}
 Based on this Poisson algebra, one can check that the discretized Gaussian constraint and edge simplicity constraint form a first class constraint system, which reads
\begin{eqnarray}
\label{firstclassalgb}
\{S_e, S_e\}\propto S_e\,,\,\, \{S_e, S_v\}\propto S_e,\,\,\{G_v, G_v\}\propto G_v,\,\,\{G_v, S_e\}\propto S_e,\,\,\{G_v, S_v\}\propto S_v, \quad b(e)=v,
\end{eqnarray}
where the brackets within $G_v\approx0$ are isomorphic to the $so(D+1)$ algebra. The Poisson brackets between the vertex simplicity constraints are the problematic ones, which give the open anomalous brackets
\begin{eqnarray}
\label{anomalousalgb}
\{S_{v,e,e'},S_{v,e,e''}\}\propto \emph{anomaly term}
\end{eqnarray}
where the $ \emph{anomaly term}$ are not proportional to any of the existing constraints in the phase space.

 The generalized twisted geometric parametrization of the discrete phase space provides us a proper treatment of the anomalous simplicity constraints. With the closure constraint, simplicity constraint and the $(D-1)$-faces' shape matching condition being solved properly, the generalized twisted geometric parameters reproduce the Regge geometry correctly \cite{PhysRevD.103.086016}. This generalized twisted geometric parametrization is given as follows. Recall the discrete phase space $\times_{e\in \gamma}T^\ast SO(D+1)_e$ associated to the given graph $\gamma$. In this space, the constraint surface defined by the edge simplicity constraint is given by
\begin{equation}
\times_{e\in \gamma}T_{\text{s}}^\ast SO(D+1)_e:=\{(h_e,X_e)\in \times_{e\in \gamma}T^\ast SO(D+1)_e|X_{e}^{[IJ}X_{e}^{KL]}=0\}.
\end{equation}
Without loss of generality, let us focus on the edge simplicity constraint surface $T_{\text{s}}^\ast SO(D+1)$ for a single edge.  It has been shown that the generalized twisted geometry variables $(V,\tilde{V},\xi^o, \eta,\bar{\xi}^\mu)\in P:=Q_{D-1}\times Q_{D-1}\times T^*S^1\times SO(D-1)$ give an angle-bivector parametrization of $T_{\text{s}}^\ast SO(D+1)$ \cite{PhysRevD.103.086016}. Let us explain these generalized twisted geometry variables briefly. The bi-vector $V$ or $\tilde{V}$ constitutes the homogeneous space $Q_{D-1}:=SO(D+1)/(SO(2)\times SO(D-1))$ with $SO(2)\times SO(D-1)$ being the maximum subgroup in $SO(D+1)$ preserving the bivector $\tau_o:=2\delta_1^{[I}\delta_2^{J]}$. The real number $\eta$ takes values in $\mathbb{R}$. The angle $\xi^o\in [-\pi,\pi)$ and the tensor $\bar{\xi}^\mu$ satisfy $e^{\xi^o\tau_o}\in SO(2)$ and $e^{\bar{\xi}^\mu\bar{\tau}_\mu}:=\bar{u}\in SO(D-1)$ with $\mu\in\{1,...,\frac{(D-1)(D-2)}{2}\}$ and $\bar{\tau}_\mu$ being a basis of $so(D-1)$. To express the intrinsic curvature by twisted geometry parameters, one need specify a pair of the $SO(D+1)$ valued Hopf sections $u(V)$ and $\tilde{u}( \tilde{V})$ satisfying $V=u(V)\tau_ou(V)^{-1}$ and $\tilde{V}=-\tilde{u}(\tilde{V})\tau_o\tilde{u}(\tilde{V})^{-1}$.  Then, with the specified $u(V)$ and $\tilde{u}( \tilde{V})$, the generalized twisted geometry parametrization associated with a single edge can be given by the map
\begin{eqnarray}\label{para}
P\ni(V,\tilde{V},\xi^o,\eta,\bar{\xi}^\mu)\mapsto(h, X)\in T_{\text{s}}^\ast SO(D+1):&& X=\frac{1}{2}\eta V=\frac{1}{2}\eta u(V)\tau_ou(V)^{-1}\\\nonumber
&&h=u(V)\,e^{\bar{\xi}^\mu\bar{\tau}_\mu}e^{\xi^o\tau_o}\,\tilde{u}(\tilde{V})^{-1}.
\end{eqnarray}
 One can see that in the image of this map, the bi-vector formulation of $X=\frac{1}{2}\eta u\tau_ou^{-1}$ always solves the edge simplicity constraint $X^{[IJ}X^{KL]}=0$. Besides,  this map is a two-to-one double covering of the image. In other words, under this map \eqref{para}, the two points $(V,\tilde{V},\xi^o, \eta,\bar{\xi}^\mu)$ and $(-V,-\tilde{V},-\xi^o,-\eta,\dot{\xi}^\mu)$ related by $e^{\dot{\xi}^\mu\bar{\tau}_{\mu}}=e^{-\pi\tau_{13}}e^{\bar{\xi}^\mu\bar{\tau}_{\mu}}e^{\pi\tau_{13}}$ with $\tau_{13}=2\delta_1^{[I}\delta_3^{J]}$ in $P$ are mapped to the same point $(h, X)\in T_{s}^\ast \!SO(D+1) $. Hence, by selecting either branch among the two signs, one can establish a bijection map in the region $|X|\neq0$, (see more details in e.g. \cite{PhysRevD.103.086016}). In addition, we get a much simpler the Poisson structures in the twisted geometry parameter space. For instance, we have the non-vanishing Poisson bracket
 \begin{equation}\label{xiN}
 \{\xi^o, \eta\}=\frac{2\kappa}{a^{D-1}},
 \end{equation}
 where $\xi^o$ and $\eta$ capture the degrees of freedom of the extrinsic and intrinsic geometries respectively.
   Recall that the phase space on the whole graph $\gamma$ is just the Cartesian product of the phase space on each edge of $\gamma$. Hence, the twisted geometry parametrization of the phase space on a single edge $e$ can be extended to that of the whole graph $\gamma$. Then, one can impose the Gaussian constraint and vertex simplicity constraint at the vertices of $\gamma$.  Based on the twisted geometry parametrization, the anomalous vertex simplicity constraint can be treated under guiding of the geometric meaning of the parameters. It has been shown that, on the constraint surface defined by both of the edge simplicity and anomalous vertex simplicity constraints in the discrete phase space, the gauge transformation induced by the edge simplicity constraint is exactly identical with the gauge transformation induced by the non-anomalous simplicity constraint in the connection phase space in the continuum limit. This result can be illustrated as
  \begin{equation}
\bar{\xi}^\mu_e\xrightarrow{\text{continuum limit}}
\bar{K}_{aIJ}
  \end{equation}
  where $\bar{\xi}^\mu_e$ captures the pure gauge degrees of freedom with respect to simplicity constraint in holonomy $h_e$.
  Thus, to solve the Gaussian and simplicity constraints in discrete phase space correctly, one should execute the symplectic reduction with respect to edge simplicity constraint and Gaussian constraint, and then solve the vertex simplicity constraint weakly. It has been shown that,  the resulting constrained twisted geometry space covers the degrees of freedom of internal and external Regge geometry on the $D$-dimensional spatial manifold $\sigma$, with the twisted geometry parameters take certain geometric meaning in Regge geometry \cite{PhysRevD.103.086016}.
\subsection{The kinematic Hilbert space in all dimensional loop quantum gravity}

The loop quantization of the connection formulation of $(1+D)$-dimensional GR leads to a Hilbert space $\mathcal{H}$, which is given by the completion of the space of cylindrical functions on the quantum configuration space. This Hilbert space $\mathcal{H}$ can be regarded as a union of the spaces $\mathcal{H}_\gamma=L^2((SO(D+1))^{|E(\gamma)|},d\mu_{\text{Haar}}^{|E(\gamma)|})$ on all possible graphs $\gamma$,  where $E(\gamma)$ denotes the set of edges of $\gamma$ and $d\mu_{\text{Haar}}^{|E(\gamma)|}$ denotes the product of the Haar measure on $SO(D+1)$. The space $\mathcal{H}_\gamma$ is also regarded as the Hilbert space given by quantizing the discrete phase space $\times_{e\in\gamma}T^\ast SO(D+1)_e$ aforementioned. A basis of $\mathcal{H}_\gamma$ are composed of the spin-network states constructed on $\gamma$, which are given by assigning an $SO(D+1)$ representation $\Lambda$ to each edge $e\in\gamma$, and an intertwiner $i_v$ to each vertex $v\in\gamma$. More explicitly, a basis state $\Psi_{\gamma,{\vec{\Lambda}}, \vec{i}}(\vec{h}(A))$ as a wave function on $\times_{e\in\gamma}SO(D+1)_e$ can be expressed as
\begin{eqnarray}
\Psi_{\gamma,{\vec{\Lambda}}, \vec{i}}(\vec{h}(A))\equiv \bigotimes_{v\in\gamma}{i_v}\,\, \rhd\,\, \bigotimes_{e\in\gamma} \pi_{\Lambda_e}(h_{e}(A)),
\end{eqnarray}
where $\vec{h}(A):=(...,h_e(A),...), \vec{\Lambda}:=(...,\lambda_e,...)$ with $ e\in\gamma$ and $\vec{i}:=(...,i_v,...)$ with $v\in\gamma$. $\pi_{\Lambda_e}(h_{e})$ is the matrix of the holonomy $h_e$ in the representation labelled by $\Lambda_e$, and $\rhd$ represents the contraction of the intertwiners with the matrixes of holonomies. Thus, the spin-network function is simply the product of the specified matrix elements function of the holonomy, which is selected by the intertwiners at the vertices. The basic variables---holonomy and flux---can be promoted as operators in the Hilbert space $\mathcal{H}$. The resulting holonomy operator and flux operator act on the spin-network functions as
\begin{eqnarray}
 \hat{ h}_{e}(A) \Psi_{\gamma,{\vec{\Lambda}}, \vec{i}}(\vec{h}(A)) &=& { h}_e(A) \Psi_{\gamma,{\vec{\Lambda}}, \vec{i}}(\vec{h}(A)) \\\nonumber
  \hat{F}_e^{IJ}\Psi_{\gamma,{\vec{\Lambda}}, \vec{i}}(\vec{h}(A)) &=&-\mathbf{i}\hbar\kappa\beta R_e^{IJ}\Psi_{\gamma,{\vec{\Lambda}}, \vec{i}}(\vec{h}(A))
\end{eqnarray}
where $R_{e}^{IJ}:=\text{tr}((\tau^{IJ}h_e)^{\text{T}}\frac{\partial}{\partial h_e})$  is the right
invariant vector fields on $SO(D+1)$ associated to the edge $e$ and $\text{T}$ representing the transposition of the matrix.

 The kinematic physical Hilbert space can be obtained by solving the Gaussian and simplicity constraints in the Hilbert space $\mathcal{H}$. As we mentioned before, the edge simplicity constraint and Gaussian constraint should be imposed strongly. This operation results the space spanned by the edge-simple and gauge invariant spin-network states, which are labelled by the simple representations of $SO(D+1)$ at edges and the gauge invariant intertwiners at vertices. Then, the anomalous vertex simplicity constraints should be imposed weakly. The resulting weak solutions are given by the spin-network states labelled by the simple coherent intertwiners at vertices \cite{long2019coherent}. A typical spin-network state $\Psi_{\gamma,\vec{N},\vec{\mathcal{I}}^{\text{s.c.}}}(h_e(A))$ which is labelled by the gauge invariant simple coherent intertwiners $\mathcal{I}_v^{\text{s.c.}}$ at $v\in\gamma$ takes the form
\begin{equation}
\Psi_{\gamma,\vec{N},\vec{\mathcal{I}}^{\text{s.c.}}}(\vec{h}(A))=\text{tr}(\otimes_{e\in\gamma} \pi_{N_e}(h_e(A))\otimes_{v\in\gamma}\mathcal{I}_v^{\text{s.c.}})
\end{equation}
with $\pi_{N_e}(h_e(A))$ denoting the matrix of $h_e(A)$ in the simple representation labelled by an non-negative integer $N_e$, and $\mathcal{I}_v^{\text{s.c.}}$ being a so-called gauge invariant simple coherent intertwiner.
\subsection{Perelomov type coherent state of $SO(D+1)$ and coherent intertwiner}
In order to introduce the explicit definition of simple coherent intertwiners,  we need to give some basic concepts of the homogeneous harmonic functions on the $D$-sphere ($S^D$).
 The space of homogeneous harmonic functions with degree $N$ on the $D$-sphere denoted by $\mathfrak{H}_{D+1}^{N}$ is an irreducible representation space of $SO(D+1)$, with the dimensionality being given by $\dim(\mathfrak{H}_{D+1}^{N})=\dim(\pi_N)=\frac{(D+N-2)!(2N+D-1)}{(D-1)!N!}$.  In order to give an orthonormal basis of the space $\mathfrak{H}_{D+1}^{N}$, let us introduce a subgroup series $SO(D+1)\supset SO(D)\supset SO(D-1)\supset ... \supset SO(2)_{\delta_1^{[I}\delta_2^{J]}}$, with $SO(2)_{\delta_1^{[I}\delta_2^{J]}}$ being the one-parameter subgroup of $SO(D+1)$ generated by $\tau_o:=2\delta_1^{[I}\delta_2^{J]}$. Then, an orthonormal basis of the space $\mathfrak{H}_{D+1}^{N}$ is given by the set of such kind of homogenous and harmonic functions $\Xi_{D+1}^{N,\mathbf{M}}(\bm{x})$ on $S^D$, or equivalently,  denoted by $\ket{N,\mathbf{M}}$ in Dirac bracket notation, where $\mathbf{M}:=M_1,M_2,...,M_{D-1}$ with $N\geq M_1 \geq M_2\geq...\geq|M_{D-1}|$, and $N,M_1,... M_{D-2}\in\mathbb{N}$, $M_{D-1}\in \mathbb{Z}$. The labels $N, \mathbf{M}$ of the function $\Xi_{D+1}^{N,\mathbf{M}}(\bm{x})$ take the meaning that $\Xi_{D+1}^{N,\mathbf{M}}(\bm{x})$ belongs to the series of spaces $\mathfrak{H}_{2}^{M_{D-1}}\subset \mathfrak{H}_{3}^{M_{D-2}}\subset...\subset\mathfrak{H}_{D}^{M_{1}}\subset\mathfrak{H}_{D+1}^{N}$, which are the irreducible representation spaces labeled by $M_{D-1},...,M_2,M_1, N$ of the series of subgroups $SO(2)_{\delta_1^{[I}\delta_2^{J]}} \subset SO(3)\subset ... \subset SO(D)\subset SO(D+1)$ respectively \cite{vilenkin2013representation}. Based on these conventions, one can immediately gives the inner product between these harmonic functions as
\begin{equation}
\langle N,\mathbf{M}|N,\mathbf{M}'\rangle:=\int_{S^D} d\bm{x} \, \overline{\Xi_{D+1}^{N,\mathbf{M}}(\bm{x})}\Xi_{D+1}^{N,\mathbf{M}'}(\bm{x})=\delta_{\mathbf{M},\mathbf{M}'},
\end{equation}
where $d\bm{x}$ represents the normalized invariant measure on $S^D$, and $\delta_{\mathbf{M},\mathbf{M}'}=1$ if $\mathbf{M}=\mathbf{M}'$ and zero otherwise. An element $g\in SO(D+1)$ acts on a spherical harmonic function $f(\bm{x})$ on $S^D$ as
  \begin{equation}\label{gfx}
  g\circ f(\bm{x})=f(g^{-1}\circ\bm{x}).
  \end{equation}
 Correspondingly, the basis $\{\tau_{IJ}\}$ of $so(D+1)$, given by $(\tau_{IJ})^{\text{def.}}:=2\delta_I^{[K}\delta_J^{L]}$ in the definition representation space of $SO(D+1)$, are operators in $\mathfrak{H}_{D+1}^{N}$ and they act on the spherical harmonic function as
 \begin{equation}
 \tau_{IJ}\circ f(\bm{x}):=\frac{d}{dt}f(e^{-t\tau_{IJ}}\circ\bm{x})|_{t=0}
 \end{equation}
 which gives a representation of the Lie algebra
 \begin{equation}\label{[JJ]}
[\tau_{IJ},\tau_{KL}]=\delta_{IL}\tau_{JK}+\delta_{JK}\tau_{ IL}-\delta_{IK}\tau_{JL}-\delta_{JL }\tau_{IK}.
\end{equation}

Following the standard procedures introduced in \cite{GeneralizedCoherentStates}, we can construct the Perelomov type coherent states in the space $\mathfrak{H}_{D+1}^{N}$. First, the state in $\mathfrak{H}_{D+1}^{N}$ corresponding to the highest weight vector is $|N_e,\mathbf{N}_{e}\rangle$ with $\mathbf{N}_{e}=\mathbf{M}|_{M_1=...=M_{D-1}=N_{e}}$. Then, the Perelomov type coherent states in this space can be defined as $|N,V\rangle:=u(V)|N,\mathbf{N}\rangle$,  where $u(V)$ is a specific $SO(D+1)$ valued function of $V$ satisfying $V=u(V)\tau_ou(V)^{-1}$ \cite{GeneralizedCoherentStates}. Similarly, the Perelomov type coherent state $|N,V\rangle$ can be defined as $|N,V\rangle:=u(-V)|N,\bar{\mathbf{N}}\rangle$ based on the state $|N_e,\bar{\mathbf{N}}_{e}\rangle\in \mathfrak{H}_{D+1}^{N}$ corresponding to the lowest weight vector, where $\mathbf{N}_{e}=\mathbf{M}|_{M_1=...=M_{D-2}=N_{e}, M_{D-1}=-N_{e}}$. It has been shown that the Perelomov type coherent state of $SO(D+1)$ has well-behaved peakedness property for the operators $\tau_{IJ}$. For instance, the Perelomov type coherent state $|N, V\rangle$ minimizes the uncertainty of the expectation value $\langle N,V|\tau_{IJ}|N,V\rangle=\mathbf{i}N V_{IJ}$ and the Heisenberg uncertainty relation of the operators $\tau_{IJ}$. In other words, the inequality
      \begin{equation}
      \left(\bigtriangleup\!\vev{\tau_{IJ}}\right)^2 \left(\bigtriangleup\!\vev{\tau_{KL}}\right)^2\ \geq\ \frac{1}{4}\left|\vev{[\tau_{IJ},\tau_{KL}]}\right|^2
      \end{equation}
      is saturated for the state $|N, V\rangle$, where we used the abbreviation  $\vev{\hat{O}}\equiv\langle N,V|\hat{O}|N,V\rangle$ and $\bigtriangleup\!\vev{\hat{O}}\equiv\sqrt{\vev{\hat{O}^2}-\vev{\hat{O}}^2}$. In addition, the system of the Perelomov type coherent states $\{|N, V\rangle\}$ provides an over-complete basis of $\mathfrak{H}_{D+1}^N$ and the resolution of identity is given by
\begin{equation}\label{iden1}
  \dim\left(\mathfrak{H}_{D+1}^N\right)\int_{Q_{D-1}}dV|N,V\rangle\langle N,V|=\mathbb{I}_{\mathfrak{H}_{D+1}^N},
\end{equation}
where $dV$ is the invariant measure induce by the Haar measure of $SO(D+1)$ so that $\int_{Q_{D-1}}dV=1$. Another property of this kind of coherent state which will be concerned in this paper is that the coherent states $|N, V\rangle$ and $|N, V'\rangle$ with $V\neq V'$ are not mutually orthogonal unless $[V^{IJ}\tau_{IJ},V'^{KL}\tau_{KL}]=0$. This means
\begin{equation}
0\leq|\langle N,V|N, V'\rangle|\leq1
\end{equation}
with $\langle N,V|N, V'\rangle=1$ if $V=V'$, $\langle N,V|N, V'\rangle=0$ if $[V^{IJ}\tau_{IJ},V'^{KL}\tau_{KL}]=0 \ \text{and}\ V\neq V'$. Moreover, it is shown that
\begin{equation}
\langle N_e, V'_e|N_e, V_e\rangle=(\text{Ang}[V'_e, V_e])^{N_e}e^{\mathbf{i}N_e\varphi(V'_e, V_e)},
\end{equation}
where $\text{Ang}[V'_e, V_e]$ and $\varphi(V'_e, V_e)$ given by the relative angles between $V'_e$ and $V_e$ are independent of $N$ with $0\leq\text{Ang}[V'_e, V_e]\leq1$.  Especially, we have $V'_e=V_e$ if and only if $\text{Ang}[V'_e, V_e]=1$, and $\varphi(V'_e, V_e)=0$ if $\text{Ang}[V'_e, V_e]=1$.

Now we can give the explicit expressions of the coherent intertwiners constructed by the Perelomov type coherent state of $SO(D+1)$. In order to simplify the expressions, we re-orient the edges linked to $v$ to be outgoing for the graph $\gamma$ without loss of generality. Then, the gauge fixed coherent intertwiners can be defined as the tensor product $\check{\mathcal{I}}_v^{\text{c.}}(\vec{N},\vec{V}):=\otimes_{e: b(e)=v}\langle N_e,V_e|$ in the space $\mathcal{H}^{\vec{N}_e}_v:=\otimes_{ b(e)=v}\overline{\mathfrak{H}}^{N_{e},D+1}$, where $\overline{\mathfrak{H}}^{N_{e},D+1}$ is the dual space of ${\mathfrak{H}}^{N_{e},D+1}$, and $|N_e,V_e\rangle$ is the Perelomov type coherent state of $SO(D+1)$. Taking advantage of the expression of $\check{\mathcal {I}}_v^{\text{c.}}$, we refer to its group averaging as $\mathcal {I}_v^{\text{c.}}$,
 namely $\mathcal{I}_v^{\text{c.}}(\vec{N},\vec{V}):=\int_{SO(D+1)}dg\otimes_{e: b(e)=v}\langle N_e,V_e|g$. Then, the so-called simple coherent intertwiners $\check{\mathcal{I}}_v^{\text{s.c.}}$ can be defined by setting $V_e^{[IJ}V_{e'}^{KL]}=0$ with $b(e)=b(e')=v$ in their definition, and the gauge invariant version $\mathcal{I}_v^{\text{s.c.}}$ can be defined analogously. It turns out that the simple coherent intertwiners weakly solve the vertex simplicity constraint by vanishing the expectation value of vertex simplicity constraint operator. Besides, it has been shown that the simple coherent intertwiners capture correct classical spatial geometric degrees of freedom in large $N$ limit. Thus, the weak imposition in the coherent intertwiner space provides a proper treatment for the anomalous quantum vertex simplicity constraint \cite{long2019coherent}.
\section{Heat-kernel coherent state and superposition type coherent state in all dimensional loop quantum gravity}\label{sec3}
The heat-kernel coherent states in $\mathcal{H}_\gamma$ are given as the product of heat-kernel coherent states associated to each edge $e\in \gamma$, which are just the heat-kernel coherent states of $SO(D+1)$ labeled by points in the phase space $T^\ast SO(D+1)$. Generally, the heat-kernel coherent states of $SO(D+1)$ takes the form
\begin{equation}\label{eq:kt}
  K_t(h,H)=\sum_{\Lambda}\dim(\pi_{\Lambda})e^{t\Delta}\chi^{\pi_{\Lambda}}(hH^{-1}),
\end{equation}
where $H\in SO(D+1)_{\mathbb{C}}\cong T^\ast SO(D+1)$, $\pi_{\Lambda}$ denotes the representation of $SO(D+1)$ labelled by $\Lambda$, $-\Delta$ is the Casimir operator of $SO(D+1)$, and $\chi^{\pi_{\Lambda}}(hH)$ is the trace of $hH$ in the representation $\pi_{\Lambda}$. The complicated structure of $SO(D+1)_{\mathbb{C}}$ makes a huge obstacle for studying the properties of the heat-kernel coherent state. Fortunately, in all dimensional LQG, the simplicity constraint eliminates the degrees of freedoms so that the heat-kernel coherent state of $SO(D+1)$ can be simplified when it is applied in all dimensional LQG. First, one can restrict the representations of holonomies to be the simple ones to solve the edge simplicity constraint, which means that those terms in \eqref{eq:kt} corresponding to the non-simple representations can be thrown away. Second, the physical degrees of freedom in the phase space are only contained in the simplicity constraint surface, so the labelling $H$ of heat-kernel coherent states can be restricted to be $H^o$ which takes values in the edge simple constraint surface $SO(D+1)^{\text{s.}}_{\mathbb{C}}\cong T^\ast_{\text{s.}}SO(D+1)$. Then, this procedure gives us the simple heat-kernel coherent states of $SO(D+1)$ as,
 \begin{equation}
  K_t(h,H^o)=\sum_{N}\dim(\pi_{N})e^{-N(N+D-1)t}\chi^{\pi_{N}}(h{H^o}^{-1}),
\end{equation}
where $\pi_{N}$ denotes the simple representation of $SO(D+1)$ labelled by the non-negative integer $N$.

The analysis of this simple heat-kernel coherent state follows a decomposition of the element $H^o\in T^\ast_{\text{s.}}SO(D+1)$.
Following the polar decomposition of $SO(D+1)_{\mathbb{C}}$, an element $H^o\in SO(D+1)^{\text{s.}}_{\mathbb{C}}$ can be rewritten as
\begin{equation}\label{pd}
H^o=g\exp{\left(\mathbf{i}\eta\tau_o\right)}\tilde{g}^{-1},
\end{equation}
where $\eta$ is a positive real number, $g$ and $\tilde{g}$ are two independent $SO(D+1)$ group elements. Further,
let us choose and fix two Hopf sections $u(V):\ V\mapsto u(V)\in SO(D+1)$ and $\tilde{u}(\tilde{V}):\ \tilde{V}\mapsto \tilde{u}(\tilde{V})\in SO(D+1)$. Then, an arbitrary element $g\in SO(D+1)$ or $\tilde{g}\in SO(D+1)$ can be uniquely decomposed as
\begin{equation}
g=u(V)e^{\phi\tau_o}\bar{g}\ \text{or}\  \tilde{g}=\tilde{u}(\tilde{V})e^{\tilde{\phi}\tau_o}\tilde{\bar{g}}
\end{equation}
with an angle $\phi$ or $\tilde{\phi}$, an element $\bar{g}$ or  $\tilde{\bar{g}}$ of $SO(D-1)$ preserving  $\tau_o$ and an unit bi-vector $V\in Q_{D-1}$ or $\tilde{V}\in Q_{D-1}$ satisfying $V=u(V)\tau_ou^{-1}(V)$ or $\tilde{V}=-\tilde{u}(\tilde{V})\tau_o\tilde{u}^{-1}(\tilde{V})$.
Based on these expressions, $H^o$ is finally decomposed as
\begin{equation}
H^o=u(V)e^{\phi\tau_o}\bar{g}\exp{\left(\mathbf{i}\eta\tau_o\right)} e^{-\tilde{\phi}\tau_o}\bar{\tilde{g}}^{-1}\tilde{u}^{-1}(\tilde{V})
=u(V)\bar{g} \bar{\tilde{g}}^{-1}\exp{\left(z\tau_o\right)} \tilde{u}^{-1}(\tilde{V}),
\end{equation}
where $z=(\phi-\tilde{\phi})+\mathbf{i}\eta=:\xi^o+\mathbf{i}\eta$, $\bar{g}, \bar{\tilde{g}}\in SO(D-1)$, $u(V), \tilde{u}(\tilde{V}) \in Q_{D-1}$. It is easy to see that this decomposition recovers the twisted geometry parametrization of $T^\ast_{\text{s}} SO(D+1)$ by $(\eta, V,\tilde{V},\xi^o,\bar{\xi}^\mu)$ introduced in section \ref{sec2}, with $\bar{g} \bar{\tilde{g}}^{-1}=e^{\bar{\xi}^\mu\bar{\tau}_\mu}$.

Making use of this decomposition, one can consider the large $\eta_e$ limit of the $SO(D+1)$ heat-kernel coherent state $  K_t(h_e,H_e^o)$  constructed for a given edge $e\in\gamma$. Let us focus on the curious cases with $\eta_{e}\gg1$. In this case, by choosing a proper basis, the matrix of $\exp{\left(-z_{e}\tau_o\right)}$ appearing in the decomposition of ${H_e^o}^{-1}$ can be simplified as
\begin{equation}\label{etagg1}
\langle N_{e},\mathbf{M}|\exp{\left(-z_{e}\tau_o\right)}|N_{e},\mathbf{M}'\rangle =\delta^{\mathbf{M}}_{ \ \mathbf{M}'}e^{-\mathbf{i}z_{e}M_{D-1}} =\delta^{\mathbf{M}}_{\ \mathbf{M}'}\exp{\left(\eta_{e}N_{e}\right)} \left(\delta_{\mathbf{M}, \mathbf{N}_{e}}e^{-\mathbf{i}\xi^o_{e}N_{e}}+\mathcal{O}(e^{-\eta_{e}})\right),
\end{equation}
where $\mathbf{N}_{e}=\mathbf{M}|_{M_1=...=M_{D-1}=N_{e}}$. Hence, we get the approximation
\begin{equation}\label{etagg2}
\sum_{\mathbf{M},\mathbf{M}'}|N_{e},\mathbf{M}\rangle\langle N_{e},\mathbf{M}|\exp{\left(-z_{e}\tau_o\right)}|N_{e},\mathbf{M}'\rangle\langle N_{e},\mathbf{M}'| \approx e^{\eta_{e}N_{e}} e^{{-\mathbf{i}}\xi^o_{e}N_{e}}|N_{e},\mathbf{N}_{e}\rangle\langle N_{e},\mathbf{N}_{e}|.
\end{equation}
Now let us insert Eq.\eqref{etagg2} into ${H_e^o}^{-1}$. Notice that $\bar{g}_{e} \bar{\tilde{g}}^{-1}_{e}$ fixes  $|N_{e},\mathbf{N}_{e}\rangle$ as $\bar{g}_{e} \bar{\tilde{g}}^{-1}_{e}|N_{e},\mathbf{N}_{e}\rangle=|N_{e},\mathbf{N}_{e}\rangle$. Then we have
 \begin{equation}
  K_t(h_e,H_e^o)\stackrel{\text{large}\  \eta_e}{=}\tilde{\Psi}_{\mathbb{H}^o_e}(h_e)=\sum_{N_e}\dim(\pi_{N_e}) e^{-tN_e(N_e+D-1)}(e^{(\eta_e -\mathbf{i}\xi^o_e)N_e}\langle N_e,\mathbf{N}|u_e^{-1}h_e\tilde{u}_e|N_e,\mathbf{N}\rangle,
 \end{equation}
 where we define $\mathbb{H}^o_e:=(\eta_e, \xi^o_e, V_e, \tilde{V}_e)$.  In fact, the state $\tilde{\Psi}_{\mathbb{H}^o_e}(h_e)$ is just the superposition type coherent state on an edge $e$ in all dimensional LQG \cite{Long:2021xjm}. It has been shown that the superposition type coherent state $\tilde{\Psi}_{\gamma,\vec{\mathbb{H}}^o}$ on the graph $\gamma$ in all dimensional LQG  provides a resolution of identity of the space $\mathcal{H}_{\gamma}$ if the range of labelling $\eta_e$ is extended to be $\mathbb{R}$. Additionally, the peakedness property of this coherent states is studied based on the simplest one loop graph \cite{Long:2021xjm}.

 In general, the superposition type coherent state is given by selecting the terms corresponding to the highest weight vector of representation of $SO(D+1)$ in the simple heat-kernel coherent state. These terms give the superpositions over quantum numbers and holonomy matrix element selected by the Perelomov type coherent state of $SO(D+1)$. The peakedness property of the resulting superposition type coherent state is ensured by the Gaussian superposition and the well-behaved peakedness property of the Perelomov type coherent state of $SO(D+1)$. One should note that the terms corresponding to the highest weight vector of the representation of $SO(D+1)$ in the simple heat-kernel coherent state is relatively large in large $\eta_e$ limit, so it also dominates the property of heat-kernel coherent state. Besides, the labelling of the superposition type coherent state excludes the dependence on the gauge component $\bar{g}_{e} \bar{\tilde{g}}^{-1}_{e}$ appearing in the labelling of simple heat-kernel coherent state, which implies that the superposition type coherent state is more physically reasonable at semiclassical level.

 In fact, there is another family of Perelomov type coherent state of $SO(D+1)$ given by the lowest weight vector of representation of $SO(D+1)$. Thus, there is another possibilities to construct the coherent state in all dimensional LQG by modifying the simple heat-kernel coherent state of $SO(D+1)$. A new proposal of coherent state in all dimensional LQG called the twisted geometry coherent state will be considered in the following sections, which contains the terms corresponding to both the highest and lowest weight vector of representation of $SO(D+1)$ in the simple heat-kernel coherent state. In the following part of this paper, we will study the basic properties of the twisted geometry coherent state and compare it with the previous ones.
\section{Generalized twisted geometry coherent states in all dimensional loop quantum gravity}\label{sec4}
Inspired by the twisted geometric coherent state in the (1+3)-dimensional $SU(2)$ LQG \cite{Calcinari_2020}, we propose the generalized twisted geometry coherent states in all dimensional LQG, which reads
\begin{eqnarray}\label{TGCS}
\breve{\Psi}_{\gamma,\vec{\mathbb{H}}^o_e}(\vec{h}_e)&:=&\prod_e\sum_{N_e}(\dim(\pi_{N_e}))^{3/2} e^{-tN_e(N_e+D-1)}(e^{(\eta_e -\mathbf{i}\xi^o_e)(N_e+\frac{D-1}{2})}\langle N_e,\mathbf{N}|u_e^{-1}h_e\tilde{u}_e|N_e,\mathbf{N}\rangle\\\nonumber
&&+e^{(-\eta_e +\mathbf{i}\xi^o_e)(N_e+\frac{D-1}{2})}\langle N_e,\bar{\mathbf{N}}|u_e^{-1}h_e\tilde{u}_e|N_e,\bar{\mathbf{N}}\rangle).
\end{eqnarray}
This coherent state can also be rewritten as
\begin{eqnarray}\label{twcs}
\nonumber\breve{\Psi}_{\gamma,\vec{\mathbb{H}}^o_e}(\vec{h}_e)&:=&\prod_e\sum_{N_e}(\dim(\pi_{N_e}))^{3/2} e^{\frac{(\eta_e)^2+t^2(D-1)^2}{4t}}\left(\exp(-t(\frac{\eta_e}{2t}-d_{N_e})^2)e^{-\mathbf{i}\xi^o_ed_{N_e}}\langle N_e,\mathbf{N}|u_e^{-1}h_e\tilde{u}_e|N_e,\mathbf{N}\rangle\right.\\
&&+\left.\exp(-t(\frac{\eta_e}{2t}+d_{N_e})^2)e^{\mathbf{i}\xi^o_ed_{N_e}}\langle N_e,\bar{\mathbf{N}}|u_e^{-1}h_e\tilde{u}_e|N_e,\bar{\mathbf{N}}\rangle\right)
\end{eqnarray}
 where $d_{N_e}\equiv (N_e+\frac{D-1}{2})$. The second term in the bracket in \eqref{twcs} is exponentially suppressed for $t\rightarrow0$ and large $\eta_e$. Then the superposition type coherent state can be recovered up to some prefactors.

The reason for the presence of the both highest and lowest weights in \eqref{TGCS} is the $\mathbb{Z}_2$ symmetry of the twisted geometry parametrization of $T^\ast_{\text{s}}SO(D+1)$, and the existence of two families of Perelomov type coherent states for $SO(D+1)$. It is easy to check that Eq.\eqref{twcs} is invariant under the transformation  $(V,\tilde{V},\xi^o, \eta)$ and $(-V,-\tilde{V},-\xi^o,-\eta)$. In next subsection, we will show that both of the terms with the highest and lowest weights are necessary, in order to proof that the coherent states are able to provide a resolution of identity with $\eta_e\in\mathbb{R}_+$.
\subsection{Resolution of the identity}
Let us denoted by $\mathcal{H}^{\text{s.}}_\gamma$ the space spanned by the spin-network functions constructed on $\gamma$,  with their edges being labelled by simple representations.
With the twisted geometry coherent state, the resolution of the identity in $\mathcal{H}^{\text{s.}}_\gamma$ reads
\begin{equation}\label{resoid}
\mathbbm{1}_{\mathcal{H}^{\text{s.}}_\gamma}=
\int_{\mathbb{R}_+^{|E(\gamma)|}}(\prod_{e\in\gamma}\frac{d\eta_e}{\sqrt{2\pi t}}e^{-\frac{\eta_e^2+t^2(D-1)^2}{2t}}) \int_{\times_{v}\mathcal{P}_v}(\prod_{v\in \gamma}d\vec{V}_{e_v})\int_{\times_{e}S_e^1}(\prod_{e\in\gamma}\frac{d\xi^o_{e}}{2\pi})|\breve{\Psi}_{\gamma,\vec{\mathbb{H}}^o_e}\rangle\langle \breve{\Psi}_{\gamma,\vec{\mathbb{H}}^o_e}|, 
\end{equation}
where $d\eta_e$ is the Lebesgue measure on $\mathbb{R}$, $|E(\gamma)|$ represents the number of the edges of $\gamma$, and $d\vec{V}_{e_v}:=\prod_{e: b(e)=v}dV_{e}$ with $\int_{\times_{v} \mathcal{P}_v}\prod_{v}d\vec{V}_{e_v}=1$
 by their definitions. Here it should be reminded that we re-orient the edges
 to be outgoing at each $v$ for the convenience of specific expression.
\\
\\
\textbf{Proof:} We first notice the following inner product
 \begin{equation}
 \langle\gamma, \vec{N}''_{e},\vec{{\mathcal{I}}}''_v|\gamma, \vec{N}'_{e},\vec{{\mathcal{I}}}'_v\rangle= \delta_{(\vec{N}',\vec{N}'')} \frac{1 }{\prod_{e\in\gamma}\dim{(\pi_{N'_e})}} \prod_{v\in\gamma}\langle{{\mathcal{I}}}''_{\vec{N}'_{e_v}}|{{\mathcal{I}}}'_{\vec{N}'_{e_v}}\rangle
 \end{equation}
 between the spin-network functions in $\mathcal{H}^{\text{s.}}_\gamma$, where $|\gamma, \vec{N}'_{e},\vec{{\mathcal{I}}}'_v\rangle$  represents the spin-network function $\Psi_{\gamma, \vec{N}'_{e},\vec{{\mathcal{I}}}'_v}(\vec{h}(A))$, and ${{\mathcal{I}}}'_v={{\mathcal{I}}}'_{\vec{N}'_{e_v}}$ is an element in the intertwiner space $\mathcal{H}_v^{\vec{N}_e}$ at $v$. Then we can verify that Eq.\eqref{resoid} is a resolution of the identity of the state space $\mathcal{H}^{\text{s.}}_\gamma$ via
\begin{eqnarray}
\nonumber&&\int_{\mathbb{R}_+^{|E(\gamma)|}}(\prod_{e\in\gamma}\frac{d\eta_e}{\sqrt{2\pi t}}e^{-\frac{\eta_e^2+t^2(D-1)^2}{2t}}) \int_{\times_{v}\mathcal{P}_v}(\prod_{v\in \gamma}d\vec{V}_{e_v})\int_{\times_{e}S_e^1}(\prod_{e\in\gamma}\frac{d\xi^o_{e}}{2\pi})\langle\gamma, \vec{N}''_{e},\vec{{\mathcal{I}}}''_v|\breve{\Psi}_{\gamma,\vec{\mathbb{H}}^o_e}\rangle\langle \breve{\Psi}_{\gamma,\vec{\mathbb{H}}^o_e}|\gamma, \vec{N}'_{e},\vec{{\mathcal{I}}}'_v\rangle\\\nonumber
&=&\int_{\mathbb{R}_+^{|E(\gamma)|}}(\prod_{e\in\gamma}\frac{d\eta_e}{\sqrt{2\pi t}}) \int_{\times_{v}\mathcal{P}_v}(\prod_{v\in \gamma}d\vec{V}_{e_v})\int_{\times_{e}S_e^1}(\prod_{e\in\gamma}\frac{d\xi^o_{e}}{2\pi})
 \left(\prod_{e} \sqrt{\dim(\pi_{N'_e})\dim(\pi_{N''_e})}\right) \Big(\text{T}_1+\text{T}_2+\text{T}_3+\text{T}_4\Big)\\\nonumber
  &=& \delta_{(\vec{N}',\vec{N}'')} \int_{\mathbb{R}_+^{|E(\gamma)|}}(\prod_{e\in\gamma}\frac{d\eta_e}{\sqrt{2\pi t}}) \int_{\times_{v}\mathcal{P}_v}(\prod_{v\in \gamma}d\vec{V}_{e_v})(\prod_{e}\dim(\pi_{N'_e}))\\\nonumber
 && \left(\prod_{e}\exp(-2t(\frac{\eta_e}{2t}-d_{N'_e})^2) \cdot \prod_{v}\langle {{\mathcal{I}}}''_{\vec{N}'_{e_v}}|{\check{\mathcal{I}}}_v(\vec{N}'_{e_v},\vec{V}_{e_v})\rangle \langle {\check{\mathcal{I}}}_v(\vec{N}'_{e_v},\vec{V}_{e_v})| {{\mathcal{I}}}'_{\vec{N}'_{e_v}}\rangle\right. \\\nonumber
 && \left.+\prod_{e}\exp(-2t(\frac{\eta_e}{2t}+d_{N'_e})^2) \cdot \prod_{v}\langle {{\mathcal{I}}}''_{\vec{N}'_{e_v}}|{\check{\mathcal{I}}}_v(\vec{N}'_{e_v},-\vec{V}_{e_v})\rangle \langle {\check{\mathcal{I}}}_v(\vec{N}'_{e_v},-\vec{V}_{e_v})| {{\mathcal{I}}}'_{\vec{N}'_{e_v}}\rangle\right) \\\nonumber
&=& \delta_{(\vec{N}',\vec{N}'')} \frac{1 }{\prod_{e\in\gamma}\dim{(\pi_{N'_e})}} \prod_{v}\langle {{\mathcal{I}}}''_{\vec{N}'_{e_v}}| {{\mathcal{I}}}'_{\vec{N}'_{e_v}}\rangle\\\nonumber
&&\cdot \int_{\mathbb{R}_+^{|E(\gamma)|}}(\prod_{e\in\gamma}\frac{d\eta_e}{\sqrt{2\pi t}}) \left((\prod_{e}\exp(-2t(\frac{\eta_e}{2t}-d_{N'_e})^2)) +(\prod_{e}\exp(-2t(\frac{\eta_e}{2t}+d_{N'_e})^2)) \right) \\
&=& \delta_{(\vec{N}',\vec{N}'')} \frac{1 }{\prod_{e\in\gamma}\dim{(\pi_{N'_e})}} \prod_{v}\langle {{\mathcal{I}}}''_{\vec{N}'_{e_v}}| {{\mathcal{I}}}'_{\vec{N}'_{e_v}}\rangle,
\end{eqnarray}
where we used the definition of $d\vec{V}_{e_v}$
  in the third equal, and defined
\begin{eqnarray}
\text{T}_1&:=&\prod_{e}\exp(-t(\frac{\eta_e}{2t}-d_{N''_e})^2-t(\frac{\eta_e}{2t}-d_{N'_e})^2)e^{\mathbf{i}\xi^o_{e} (N'_{e}-N''_e)}\\\nonumber
&&\cdot\prod_{v}\langle {{\mathcal{I}}}''_{\vec{N}''_{e_v}}|{\check{\mathcal{I}}}_v(\vec{N}''_{e_v},\vec{V}_{e_v})\rangle \langle {\check{\mathcal{I}}}_v(\vec{N}'_{e_v},\vec{V}_{e_v})| {{\mathcal{I}}}'_{\vec{N}'_{e_v}}\rangle\\\nonumber
\text{T}_2&:=&\prod_{e}\exp(-t(\frac{\eta_e}{2t}+d_{N''_e})^2-t(\frac{\eta_e}{2t}+d_{N'_e})^2) e^{\mathbf{i}\xi^o_{e}(N''_{e}-N'_e)}\\\nonumber
&&\cdot \prod_{v}\langle {{\mathcal{I}}}''_{\vec{N}''_{e_v}}|{\check{\mathcal{I}}}_v(\vec{N}''_{e_v},-\vec{V}_{e_v})\rangle \langle {\check{\mathcal{I}}}_v(\vec{N}'_{e_v},-\vec{V}_{e_v})| {{\mathcal{I}}}'_{\vec{N}'_{e_v}}\rangle
\\\nonumber
\text{T}_3&:=&\prod_{e}\exp(-t(\frac{\eta_e}{2t}-d_{N''_e})^2-t(\frac{\eta_e}{2t}+d_{N'_e})^2) e^{-\mathbf{i}\xi^o_{e}(N''_{e}+N'_e+D+1)}\\\nonumber
&&\cdot\prod_{v}\langle {{\mathcal{I}}}''_{\vec{N}''_{e_v}}|{\check{\mathcal{I}}}_v(\vec{N}''_{e_v},\vec{V}_{e_v})\rangle \langle {\check{\mathcal{I}}}_v(\vec{N}'_{e_v},-\vec{V}_{e_v})|{{\mathcal{I}}}'_{\vec{N}'_{e_v}}\rangle
\\\nonumber
\text{T}_4&:=&\prod_{e}\exp(-t(\frac{\eta_e}{2t}+d_{N''_e})^2-t(\frac{\eta_e}{2t}-d_{N'_e})^2) e^{\mathbf{i}\xi^o_{e}(N''_{e}+N'_e+D+1)}\\\nonumber
&&\prod_{v}\langle \vec{{\mathcal{I}}}''_{\vec{N}''_{e_v}}|{\check{\mathcal{I}}}_v(\vec{N}''_{e_v},-\vec{V}_{e_v})\rangle \langle {\check{\mathcal{I}}}_v(\vec{N}'_{e_v},\vec{V}_{e_v})| {{\mathcal{I}}}'_{\vec{N}'_{e_v}}\rangle.
\end{eqnarray}
  This complete the proof. \\$\square$\\

  The resolution of identity here only involves the integral of $\eta_e$ in the range $\mathbb{R}_+$, instead of the whole $\mathbb{R}$ for the superposition type coherent state. It has been discussed that the superposition type coherent states with $\eta_e<0$ are not well-behaved coherent states, even though they are still necessary to construct the resolution of identity \cite{Long:2021xjm}. However, this problem disappears here because the $\mathbb{Z}_2$ symmetry ensure that the twisted geometry coherent states with $\eta_e>0$ are able to give the resolution of identity.
  \subsection{Expectation values}
  We first calculate the norm of the states,
\begin{eqnarray}
&&||\breve{\Psi}_{\gamma,\vec{\mathbb{H}}^o_e}||^2\\\nonumber
&:=&\prod_e e^{\frac{(\eta_e)^2+t^2(D-1)^2}{2t}}\sum_{N_e}(\dim(\pi_{N_e}))^2 \left(\exp(-2t(\frac{\eta_e}{2t}-d_{N_e})^2) +\exp(-2t(\frac{\eta_e}{2t}+d_{N_e})^2)\right)\\\nonumber
&\xlongequal[t\rightarrow 0 ]{\text{large }\! \eta_e}&(\sqrt{\frac{\pi}{2t}})^{|E(\gamma)|}\prod_e e^{\frac{(\eta_e)^2+t^2(D-1)^2}{2t}}(\breve{\text{Poly}}(\frac{\eta_e}{2t}))^2,
\end{eqnarray}
where $\breve{\text{Poly}}(x)$ is a polynomial of $x$ satisfying $\breve{\text{Poly}}(N_e)=\dim(\pi_{N_e})$. In the following part of this subsection, we will consider the expectation value of the flux and holonomy operators in the generalized twisted geometry coherent state and evaluate their uncertainties.
  \subsubsection{Flux operator}
  To compute the expectation value of the fluxes, we take advantage of the property of the Perelomov type coherent states of $SO(D+1)$. Then the calculation shows that
\begin{eqnarray}
&&\langle\hat{F}_e^{IJ}\rangle_{\gamma,\vec{\mathbb{H}}^o_e} := \frac{\langle\breve{\Psi}_{\gamma,\vec{\mathbb{H}}^o_e}|\hat{F}_e^{IJ} |\breve{\Psi}_{\gamma,\vec{\mathbb{H}}^o_e}\rangle}{||\breve{\Psi}_{\gamma,\vec{\mathbb{H}}^o_e}||^2}\\\nonumber
&=&-\mathbf{i}\hbar\kappa\beta\frac{\sum_{N_e}(\dim(\pi_{N_e}))^2 \exp(-2t(\frac{\eta_e}{2t}-d_{N_e})^2)\langle N_e, V_e|\tau^{IJ}_{(N_e)}| N_e, V_e\rangle}{\sum_{N_e}(\dim(\pi_{N_e}))^2 \left(\exp(-2t(\frac{\eta_e}{2t}-d_{N_e})^2)+\exp(-2t(\frac{\eta_e}{2t}+d_{N_e})^2)\right)}\\\nonumber
&&-\mathbf{i}\hbar\kappa\beta\frac{\sum_{N_e}(\dim(\pi_{N_e}))^2 \exp(-2t(\frac{\eta_e}{2t}+d_{N_e})^2)\langle N_e, -V_e|\tau^{IJ}_{(N_e)}| N_e, -V_e\rangle}{\sum_{N_e}(\dim(\pi_{N_e}))^2 \left(\exp(-2t(\frac{\eta_e}{2t}-d_{N_e})^2)+\exp(-2t(\frac{\eta_e}{2t}+d_{N_e})^2)\right)}\\\nonumber
&=&\hbar\kappa\beta V^{IJ}\frac{\sum_{N_e}(\dim(\pi_{N_e}))^2N_e \left(\exp(-2t(\frac{\eta_e}{2t}-d_{N_e})^2)-\exp(-2t(\frac{\eta_e}{2t}+d_{N_e})^2)\right)}{\sum_{N_e}(\dim(\pi_{N_e}))^2 \left(\exp(-2t(\frac{\eta_e}{2t}-d_{N_e})^2)+\exp(-2t(\frac{\eta_e}{2t}+d_{N_e})^2)\right)}\\\nonumber
&\xlongequal[t\rightarrow 0 ]{\text{large }\! \eta_e}&(\frac{\eta_e}{2t}-\frac{D-1}{2})\hbar\kappa\beta V^{IJ}
\end{eqnarray}
where $V^{IJ}\tau_{IJ}=2u\tau_ou^{-1}$ and we neglected the terms suppressed by $e^{-\eta_e^2/t}$ and $\frac{t}{\eta_e}$. Similarly we have
\begin{eqnarray}
&&\langle\hat{F}_e^{IJ}\hat{F}_{e,IJ}\rangle_{\gamma,\vec{\mathbb{H}}^o_e} := \frac{\langle\breve{\Psi}_{\gamma,\vec{\mathbb{H}}^o_e}|\hat{F}_e^{IJ}\hat{F}_{e,IJ} |\breve{\Psi}_{\gamma,\vec{\mathbb{H}}^o_e}\rangle}{||\breve{\Psi}_{\gamma,\vec{\mathbb{H}}^o_e}||^2}\\\nonumber
&=&2(\hbar\kappa\beta)^2 \frac{\sum_{N_e}(\dim(\pi_{N_e}))^2N_e(N_e+D-1) \left(\exp(-2t(\frac{\eta_e}{2t}-d_{N_e})^2)+\exp(-2t(\frac{\eta_e}{2t}+d_{N_e})^2)\right)}{\sum_{N_e}(\dim(\pi_{N_e}))^2 \left(\exp(-2t(\frac{\eta_e}{2t}-d_{N_e})^2)+\exp(-2t(\frac{\eta_e}{2t}+d_{N_e})^2)\right)}\\\nonumber
&\xlongequal[t\rightarrow 0 ]{\text{large }\! \eta_e}&(\frac{\eta^2_e}{2t^2}-\frac{(D-1)^2}{2})(\hbar\kappa\beta)^2.
\end{eqnarray}
Then we can evaluate the relative uncertainty as
\begin{equation}
\Delta(\langle\hat{F}_e^{IJ}\rangle_{\gamma,\vec{\mathbb{H}}^o_e})=\left(\frac{\langle\hat{F}_e^{IJ}\hat{F}_{e,IJ}\rangle_{\gamma,\vec{\mathbb{H}}^o_e}}{\langle\hat{F}_e^{IJ}\rangle_{\gamma,\vec{\mathbb{H}}^o_e} \langle\hat{F}_{e,IJ}\rangle_{\gamma,\vec{\mathbb{H}}^o_e}}-1\right)\xlongequal[t\rightarrow 0 ]{\text{large }\! \eta_e}0.
\end{equation}
It is easy to conclude that the expectation value of flux operator is approximated by its classical labelling in the twisted geometry coherent state with the vanishing relative uncertainty in the limits of $t\rightarrow0$ for large $\eta_e$.
  \subsubsection{Holonomy operator}
  The holonomy operator should be considered for the cases that $(D+1)$ is even or odd separately. Let us first consider that $(D+1)$ is even.
  To give the explicit expression of the matrix element of holonomy, we need a set of bi-vectors $\{V_{\imath\jmath}=2\delta_\imath^{[I}\delta_\jmath^{J]}|(\imath,\jmath)\in \{(1,2),(2,1),(3,4),(4,3),...,(D,D+1),(D+1,D)\}\}$ in $\mathbb{R}^{D+1}$.
 For a given coherent state $\breve{\Psi}_{\gamma,\vec{\mathbb{H}}^o_e}$, let us focus on the operator $(\widehat{u_e^{-1}h_e\tilde{u}_e})_{\imath\jmath,\imath'\jmath'}$ corresponding to the matrix element $\langle 1, V_{\imath\jmath}|u_e^{-1}h_e\tilde{u}_e|1,V_{\imath'\jmath'}\rangle$ of holomomy $h_e$, where $\{|1,V_{\imath\jmath}\rangle|(\imath,\jmath)\in\{(1,2),(2,1),(3,4),(4,3),...,(D,D+1),(D+1,D)\}\}$ is an orthonormal basis of the definition representation space of $SO(D+1)$ (see details in Appendix \ref{app2}). Define
 \begin{equation}
 \langle\hat{h}_{e,\imath\jmath,\imath'\jmath'}\rangle_{\gamma,\vec{\mathbb{H}}^o_e} := \frac{\langle\breve{\Psi}_{\gamma,\vec{\mathbb{H}}^o_e}|(\widehat{u_e^{-1}h_e\tilde{u}})_{\imath\jmath,\imath'\jmath'} |\breve{\Psi}_{\gamma,\vec{\mathbb{H}}^o_e}\rangle}{||\breve{\Psi}_{\gamma,\vec{\mathbb{H}}^o_e}||^2}.
 \end{equation}
   Then the calculation shows that
\begin{eqnarray}
&&\langle\hat{h}_{e,12,12}\rangle_{\gamma,\vec{\mathbb{H}}^o_e} := \frac{\langle\breve{\Psi}_{\gamma,\vec{\mathbb{H}}^o_e}|(\widehat{u_e^{-1}h_e\tilde{u}})_{12,12} |\breve{\Psi}_{\gamma,\vec{\mathbb{H}}^o_e}\rangle}{||\breve{\Psi}_{\gamma,\vec{\mathbb{H}}^o_e}||^2}\\\nonumber
&&\xlongequal[t\rightarrow 0 ]{\text{large }\! \eta_e}\frac{\sum_{N_e\in\mathbb{N}_+}(\dim(\pi_{N_e+1}))^{1/2} \dim(\pi_{N_e}))^{3/2} \exp(-t(\frac{\eta_e}{2t}-d_{N_e})^2-t(\frac{\eta_e}{2t}-d_{N_e}-1)^2)e^{\mathbf{i}\xi^o_e} } {\sum_{N_e\in\mathbb{N}_+}(\dim(\pi_{N_e}))^2 \left(\exp(-2t(\frac{\eta_e}{2t}-d_{N_e})^2)+\exp(-2t(\frac{\eta_e}{2t}+d_{N_e})^2)\right)}\\\nonumber
&&=e^{\mathbf{i}\xi^o_e}e^{-t/2}\frac{\sum_{N_e\in\mathbb{N}_+}(\dim(\pi_{N_e+1}))^{1/2} \dim(\pi_{N_e}))^{3/2} \exp(-2t(\frac{\eta_e}{2t}-d_{N_e}-\frac{1}{2})^2) }{\sum_{N_e\in\mathbb{N}_+}(\dim(\pi_{N_e}))^2 \left(\exp(-2t(\frac{\eta_e}{2t}-d_{N_e})^2)+\exp(-2t(\frac{\eta_e}{2t}+d_{N_e})^2)\right)}\\\nonumber
&&\xlongequal[t\rightarrow 0 ]{\text{large }\! \eta_e}e^{\mathbf{i}\xi^o_e}e^{-t/2},
\end{eqnarray}
where we used the results from Appendix \ref{app2} in the first ``equal sign''.
Similarly, we have
\begin{eqnarray}
&&\langle\hat{h}_{e,12,21}\rangle_{\gamma,\vec{\mathbb{H}}^o_e} := \frac{\langle\breve{\Psi}_{\gamma,\vec{\mathbb{H}}^o_e}|(\widehat{u_e^{-1}h_e\tilde{u}})_{12,21} |\breve{\Psi}_{\gamma,\vec{\mathbb{H}}^o_e}\rangle}{||\breve{\Psi}_{\gamma,\vec{\mathbb{H}}^o_e}||^2}\xlongequal[t\rightarrow 0 ]{\text{large }\! \eta_e}0,
\end{eqnarray}
\begin{eqnarray}
&&\langle\hat{h}_{e,21,12}\rangle_{\gamma,\vec{\mathbb{H}}^o_e} := \frac{\langle\breve{\Psi}_{\gamma,\vec{\mathbb{H}}^o_e}|(\widehat{u_e^{-1}h_e\tilde{u}_e})_{21,12} |\breve{\Psi}_{\gamma,\vec{\mathbb{H}}^o_e}\rangle}{||\breve{\Psi}_{\gamma,\vec{\mathbb{H}}^o_e}||^2}\xlongequal[t\rightarrow 0 ]{\text{large }\! \eta_e}0,
\end{eqnarray}
and
\begin{eqnarray}
&&\langle\hat{h}_{e,21,21}\rangle_{\gamma,\vec{\mathbb{H}}^o_e} := \frac{\langle\breve{\Psi}_{\gamma,\vec{\mathbb{H}}^o_e}|(\widehat{u_e^{-1}h_e\tilde{u}})_{21,21} |\breve{\Psi}_{\gamma,\vec{\mathbb{H}}^o_e}\rangle}{||\breve{\Psi}_{\gamma,\vec{\mathbb{H}}^o_e}||^2}\\\nonumber
&&\xlongequal[t\rightarrow 0 ]{\text{large }\! \eta_e}\frac{\sum_{N_e\in\mathbb{N}_+}(\dim(\pi_{N_e-1}))^{1/2} \dim(\pi_{N_e}))^{3/2} |\alpha_1(N_e)|^2\exp(-t(\frac{\eta_e}{2t}-d_{N_e})^2-t(\frac{\eta_e}{2t}-d_{N_e}+1)^2)e^{-\mathbf{i}\xi^o_e}} {\sum_{N_e\in\mathbb{N}_+}(\dim(\pi_{N_e}))^2 \left(\exp(-2t(\frac{\eta_e}{2t}-d_{N_e})^2)+\exp(-2t(\frac{\eta_e}{2t}+d_{N_e})^2)\right)}\\\nonumber
&&=e^{-\mathbf{i}\xi^o_e}e^{-t/2}\frac{\sum_{N_e\in\mathbb{N}_+}(\dim(\pi_{N_e-1}))^{1/2} \dim(\pi_{N_e}))^{3/2} |\alpha_1(N_e)|^2\exp(-2t(\frac{\eta_e}{2t}-d_{N_e}+\frac{1}{2})^2)}{\sum_{N_e\in\mathbb{N}_+}(\dim(\pi_{N_e}))^2 \left(\exp(-2t(\frac{\eta_e}{2t}-d_{N_e})^2)+\exp(-2t(\frac{\eta_e}{2t}+d_{N_e})^2)\right)}\\\nonumber
&&\xlongequal[t\rightarrow 0 ]{\text{large }\! \eta_e}e^{-\mathbf{i}\xi^o_e}e^{-t/2},
\end{eqnarray}
where we also used the results from Appendix \ref{app2}.
Doing the same calculation, one further has
\begin{eqnarray}\label{ijij}
&&\langle\hat{h}_{e,12,\imath\jmath}\rangle_{\gamma,\vec{\mathbb{H}}^o_e}\xlongequal[t\rightarrow 0 ]{\text{large }\! \eta_e}0,\quad \langle\hat{h}_{e,21,\imath\jmath}\rangle_{\gamma,\vec{\mathbb{H}}^o_e}\xlongequal[t\rightarrow 0 ]{\text{large }\! \eta_e}0,\\\nonumber
&&\langle\hat{h}_{e,\imath\jmath,12}\rangle_{\gamma,\vec{\mathbb{H}}^o_e}\xlongequal[t\rightarrow 0 ]{\text{large }\! \eta_e}0,\quad \langle\hat{h}_{e,\imath\jmath,21}\rangle_{\gamma,\vec{\mathbb{H}}^o_e}\xlongequal[t\rightarrow 0 ]{\text{large }\! \eta_e}0,\\\nonumber
&&\langle\hat{h}_{e,\imath\jmath,\imath'\jmath'}\rangle_{\gamma,\vec{\mathbb{H}}^o_e} \xlongequal[t\rightarrow 0 ]{\text{large }\! \eta_e}0
\end{eqnarray}
for  $(\imath,\jmath)$ and $(\imath',\jmath')\in \{(3,4),(4,3),...,(D,D+1),(D+1,D)\}$.
To evaluate the uncertainty of the expectation values, we can also check that
\begin{equation}
\langle\hat{h}_{e,12,12}\hat{h}_{e,12,12}\rangle_{\gamma,\vec{\mathbb{H}}^o_e} := \frac{\langle\breve{\Psi}_{\gamma,\vec{\mathbb{H}}^o_e}|(\widehat{u_e^{-1}h_e\tilde{u}_e})_{12,12} (\widehat{u_e^{-1}h_e\tilde{u}_e})_{12,12} |\breve{\Psi}_{\gamma,\vec{\mathbb{H}}^o_e}\rangle}{||\breve{\Psi}_{\gamma,\vec{\mathbb{H}}^o_e}||^2}\xlongequal[t\rightarrow 0 ]{\text{large }\! \eta_e}e^{2\mathbf{i}\xi^o_e}e^{-2t},
\end{equation}
\begin{equation}
\langle\hat{h}_{e,21,21}\hat{h}_{e,21,21}\rangle_{\gamma,\vec{\mathbb{H}}^o_e} := \frac{\langle\breve{\Psi}_{\gamma,\vec{\mathbb{H}}^o_e}|(\widehat{u_e^{-1}h_e\tilde{u}_e})_{21,21} (\widehat{u_e^{-1}h_e\tilde{u}_e})_{21,21} |\breve{\Psi}_{\gamma,\vec{\mathbb{H}}^o_e}\rangle}{||\breve{\Psi}_{\gamma,\vec{\mathbb{H}}^o_e}||^2}\xlongequal[t\rightarrow 0 ]{\text{large }\! \eta_e}e^{-2\mathbf{i}\xi^o_e}e^{-2t}.
\end{equation}
Then, the relative uncertainty is given by
\begin{equation}
\Delta(\langle\hat{h}_{e,12,12}\rangle_{\gamma,\vec{\mathbb{H}}^o_e}) :=(\frac{\langle\hat{h}_{e,12,12}\hat{h}_{e,12,12}\rangle_{\gamma,\vec{\mathbb{H}}^o_e}}{\langle\hat{h}_{e,12,12} \rangle^2_{\gamma,\vec{\mathbb{H}}^o_e}}-1)\xlongequal[t\rightarrow 0 ]{\text{large }\! \eta_e}0,
\end{equation}
\begin{equation}
\Delta(\langle\hat{h}_{e,21,21}\rangle_{\gamma,\vec{\mathbb{H}}^o_e}) :=(\frac{\langle\hat{h}_{e,21,21}\hat{h}_{e,21,21}\rangle_{\gamma,\vec{\mathbb{H}}^o_e}}{\langle\hat{h}_{e,21,21} \rangle^2_{\gamma,\vec{\mathbb{H}}^o_e}}-1)\xlongequal[t\rightarrow 0 ]{\text{large }\! \eta_e}0.
\end{equation}
Let us turn to the case where $(D+1)$ is odd. In this case the above calculations still hold except that the equation \eqref{ijij} holds for $(\imath,\jmath)\   \text{and}\  (\imath',\jmath')\in \{(3,4),(4,3),...,(D-1,D),(D,D-1)\}$. Besides, there are extra holonomy operators $(\widehat{u_e^{-1}h_e\tilde{u}_e})_{\imath\jmath,(D+1)}$ and $(\widehat{u_e^{-1}h_e\tilde{u}_e})_{(D+1),\imath\jmath}$ with $(\imath,\jmath)\in \{(1,2),(2,1),(3,4),...,(D-1,D),(D,D-1)\}$ in this case, which are defined by
\begin{equation}
(\widehat{u_e^{-1}h_e\tilde{u}})_{\imath\jmath,(D+1)}:=\widehat{\langle1, V_{\imath\jmath}|u_e^{-1}h_e\tilde{u}_e|1,\delta_{D+1}\rangle}
\end{equation}
and
\begin{equation}
(\widehat{u_e^{-1}h_e\tilde{u}})_{(D+1),\imath\jmath}:=\widehat{\langle1,\delta_{D+1}|u_e^{-1}h_e\tilde{u}_e|1, V_{\imath\jmath}\rangle}
\end{equation}
respectively, where $|1,\delta_{D+1}\rangle$ is defined in Appendix \ref{app2}. One can also check
\begin{eqnarray}
&&\langle\hat{h}_{e,\imath\jmath,D+1}\rangle_{\gamma,\vec{\mathbb{H}}^o_e} := \frac{\langle\breve{\Psi}_{\gamma,\vec{\mathbb{H}}^o_e}|(\widehat{u_e^{-1}h_e\tilde{u}_e})_{\imath\jmath,D+1} |\breve{\Psi}_{\gamma,\vec{\mathbb{H}}^o_e}\rangle}{||\breve{\Psi}_{\gamma,\vec{\mathbb{H}}^o_e}||^2}\xlongequal[t\rightarrow 0 ]{\text{large }\! \eta_e}0,\quad\forall \ (\imath,\jmath),
\end{eqnarray}
and
\begin{eqnarray}
&&\langle\hat{h}_{e,D+1,\imath\jmath}\rangle_{\gamma,\vec{\mathbb{H}}^o_e} := \frac{\langle\breve{\Psi}_{\gamma,\vec{\mathbb{H}}^o_e}|(\widehat{u_e^{-1}h_e\tilde{u}_e})_{D+1,\imath\jmath} |\breve{\Psi}_{\gamma,\vec{\mathbb{H}}^o_e}\rangle}{||\breve{\Psi}_{\gamma,\vec{\mathbb{H}}^o_e}||^2}\xlongequal[t\rightarrow 0 ]{\text{large }\! \eta_e}0,\quad \forall \ (\imath,\jmath).
\end{eqnarray}
Note that the classical labelling $\vec{\mathbb{H}}^o_e$ of the twisted geometry coherent state gives the holonomy $h_e^o=u_ee^{\xi_e^o\tau_o}e^{\bar{\xi}_e^\mu\bar{\tau}_\mu}\tilde{u}^{-1}_e$ up to the $SO(D-1)$ element $e^{\bar{\xi}_e^\mu\bar{\tau}_\mu}$. Let us define $(u_e^{-1}h^o_e\tilde{u})_{\imath\jmath,\imath'\jmath'}=\langle 1, V_{\imath\jmath}|u_e^{-1}h^o_e\tilde{u}|1,V_{\imath'\jmath'}\rangle$. Then we have the following matrix element of $h_e^o$,
\begin{equation}\label{holoexpect1}
(u_e^{-1}h^o_e\tilde{u}_e)_{12,12}=e^{\mathbf{i}\xi^o_e},\quad (u_e^{-1}h^o_e\tilde{u}_e)_{21,21}=e^{-\mathbf{i}\xi^o_e},
\end{equation}
\begin{equation}\label{holoexpect2}
(u_e^{-1}h^o_e\tilde{u}_e)_{12,21}=(u_e^{-1}h^o_e\tilde{u}_e)_{21,12}=0,
\end{equation}
\begin{equation}\label{holoexpect3}
(u_e^{-1}h^o_e\tilde{u}_e)_{12,\imath\jmath} =(u_e^{-1}h^o_e\tilde{u}_e)_{21,\imath\jmath} =(u_e^{-1}h^o_e\tilde{u}_e)_{\imath\jmath,12}=(u_e^{-1}h^o_e\tilde{u}_e)_{\imath\jmath,21}=0, \quad \text{for}\  (\imath,\jmath)\neq (1,2)\ \text{or}\  (2,1).
\end{equation}
\begin{equation}\label{holoexpect4}
(u_e^{-1}h^o_e\tilde{u}_e)_{\imath\jmath,\imath'\jmath'}=\langle 1, V_{\imath\jmath}|e^{\bar{\xi}_e^\mu\bar{\tau}_\mu}|1,V_{\imath'\jmath'}\rangle, \quad \text{for}\  (\imath,\jmath)\neq (1,2)\ \text{or}\  (2,1).
\end{equation}
One can see that the results \eqref{holoexpect1}, \eqref{holoexpect2} and \eqref{holoexpect3} are consistent with the corresponding expectation values of the holonomy operators $(\widehat{u_e^{-1}h_e\tilde{u}_e})_{\imath\jmath,\imath'\jmath'}$ with respect to the coherent state $\breve{\Psi}_{\gamma,\vec{\mathbb{H}}^o_e}$ in the limit $t\rightarrow0$ for large $\eta_e$. Nevertheless, the result \eqref{holoexpect4} capture the gauge degrees of freedom with respect to simplicity constraint, while the corresponding expectation value of $(\widehat{u_e^{-1}h_e\tilde{u}_e})_{\imath\jmath,\imath'\jmath'}$ are vanishing so that they are independent of the gauge degrees of freedom with respect to simplicity constraint. This inconsistency shows the divergent results of the different treatments of the gauge degrees of freedom in classical and quantum theory. In other words, the gauge degrees of freedom are treated by some gauge fixing in the classical theory while they are treated by taking averaging with respect to the gauge transformation in quantum theory.
\subsection{Peakedness properties}
Notice that the twisted geometry coherent state $\breve{\Psi}_{\gamma,\vec{\mathbb{H}}^o_e}$ on $\gamma$ is the product of the twisted geometry coherent state
\begin{eqnarray}
\breve{\Psi}_{\mathbb{H}^o_e}(h_e)&=&\sum_{N_e}(\dim(\pi_{N_e}))^{3/2} e^{-tN_e(N_e+D-1)}(e^{(\eta_e -\mathbf{i}\xi^o_e)(N_e+\frac{D-1}{2})}\langle N_e,\mathbf{N}|u_e^{-1}h_e\tilde{u}_e|N_e,\mathbf{N}\rangle\\\nonumber
&&+e^{(-\eta_e +\mathbf{i}\xi^o_e)(N_e+\frac{D-1}{2})}\langle N_e,\bar{\mathbf{N}}|u_e^{-1}h_e\tilde{u}_e|N_e,\bar{\mathbf{N}}\rangle)
\end{eqnarray}
 on each edge $e\in\gamma$. Thus, in the following calculations and analysis, we can only consider the twisted geometry coherent state on a single edge $e$ without loss of generality to simplify our expressions.
\subsubsection{The Overlap Function of the coherent states}
The overlap function for these coherent states is given by
\begin{equation}
i^t(\mathbb{H}^o_e, \mathbb{H}'^o_e):=\frac{|\langle \breve{\Psi}_{\mathbb{H}^o_e},\breve{\Psi}_{\mathbb{H}'^o_e}\rangle|^2}{||\breve{\Psi}_{\mathbb{H}^o_e}||^2 ||\breve{\Psi}_{\mathbb{H}'^o_e}||^2}
\end{equation}
with
\begin{eqnarray}
&&||\breve{\Psi}_{\mathbb{H}^o_e}||^2 \xlongequal[t\rightarrow 0 ]{ \text{large }\! \eta_e}\sqrt{\frac{\pi}{2t}} e^{\frac{(\eta_e)^2+t^2(D-1)^2}{2t}}(\breve{\text{Poly}}(\frac{\eta_e}{2t}))^2
\end{eqnarray}
and
\begin{eqnarray}
\langle \breve{\Psi}_{\mathbb{H}^o_e},\breve{\Psi}_{\mathbb{H}'^o_e}\rangle
&\stackrel{t\rightarrow 0}{=}&e^{\frac{(\eta_e)^2+(\eta'_e)^2+2t^2(D-1)^2}{4t}}\sum_{N_e}(\dim(\pi_{N_e}))^2\exp(-t(\frac{\eta_e}{2t}-d_{N_e})^2 -t(\frac{\eta'_e}{2t}-d_{N_e})^2)\\\nonumber
&&\cdot e^{\mathbf{i}N_e(\xi^o_e-\xi'^o_e+\varphi(u_e,u'_e) +\varphi(\tilde{u}_e,\tilde{u}'_e))}\exp(- N_e\widetilde{\Theta}_e)
\end{eqnarray}
for large $\eta_e, \eta'_e$. Here we defined $\widetilde{\Theta}_e:=\Theta(u_e,u'_e)+\Theta(\tilde{u}_e,\tilde{u}'_e)$ and used
\begin{equation}
\langle N_e, V'_e|N_e, V_e\rangle=\exp{(-N_e\Theta(u_e,{u'}_e))}e^{\mathbf{i}N_e\varphi(u_e,u'_e)},
\end{equation}
\begin{equation}
\langle N_e, -\tilde{V}_e|N_e,-\tilde{ V}'_e\rangle=\exp{(-N_e\Theta(\tilde{u}_e,\tilde{u}'_e))}e^{\mathbf{i}N_e\varphi(\tilde{u}_e,\tilde{u}'_e)},
\end{equation}
 where we use the convention $\Theta(u_e,{u'}_e):=-\frac{\ln|\langle N_e, V'_e|N_e, V_e\rangle|}{N_e}\geq0$, $\Theta(\tilde{u}_e,\tilde{u}'_e):=-\frac{\ln|\langle N_e, -\tilde{V}_e|N_e,-\tilde{ V}'_e\rangle|}{N_e}\geq0$, $e^{\mathbf{i}N_e\varphi(u_e,u'_e)}:=\frac{\langle N_e, V'_e|N_e, V_e\rangle}{|\langle N_e, V'_e|N_e, V_e\rangle|}$, and $e^{\mathbf{i}N_e\varphi(\tilde{u}_e,\tilde{u}'_e)}:=\frac{\langle N_e, -\tilde{V}_e|N_e, -\tilde{V}'_e\rangle}{|\langle N_e, -\tilde{V}_e|N_e, -\tilde{V}'_e\rangle|}$. We have the relations that $\Theta(u_e,u'_e)=0, \Theta(\tilde{u}_e,\tilde{u}'_e)=0$ if and only if $V_e=V'_e$, $\tilde{V}_e=\tilde{ V}'_e$ respectively, and $\varphi(u_e,u'_e)=0, \varphi(\tilde{u}_e,\tilde{u}'_e)=0$ if $V_e=V'_e$, $\tilde{V}_e=\tilde{ V}'_e$ respectively. Now let us consider the cases for $\widetilde{\Theta}_e\ll \eta_e+\eta'_e$ and $\widetilde{\Theta}_e\simeq\eta_e+\eta'_e$ or $\widetilde{\Theta}_e\gg \eta_e+\eta'_e$ respectively.

For the case $\widetilde{\Theta}_e\ll \eta_e+\eta'_e$, one can simplify the expression as follows,
\begin{eqnarray}\label{overinner}
&&\langle \breve{\Psi}_{\mathbb{H}^o_e},\breve{\Psi}_{\mathbb{H}'^o_e}\rangle\\\nonumber
&\stackrel{t\rightarrow 0}{=}&e^{\frac{(\eta_e)^2+(\eta'_e)^2+2t^2(D-1)^2}{4t}}\sum_{N_e}(\dim(\pi_{N_e}))^2\\\nonumber
&&\cdot\left(\exp(-t(\frac{\eta_e}{2t}-d_{N_e})^2 -t(\frac{\eta'_e}{2t}-d_{N_e})^2)e^{\mathbf{i}N_e(\xi^o_e-\xi'^o_e+\varphi(u_e,u'_e) +\varphi(\tilde{u}_e,\tilde{u}'_e))}\exp(- N_e\widetilde{\Theta}_e)\right)\\\nonumber
&=&e^{\frac{(\eta_e)^2+(\eta'_e)^2+2t^2(D-1)^2}{4t}}e^{-t(\frac{\eta'_e}{2t}-\frac{\eta_e}{2t})^2 +2t(\frac{\eta'_e}{4t}-\frac{\eta_e}{4t}-\frac{\widetilde{\Theta}_e}{4t})^2} e^{\mathbf{i}(\frac{\eta_e}{4t}-\frac{D-1}{2}+\frac{\eta'_e}{4t}-\frac{\widetilde{\Theta}_e}{4t})(\xi^o_e- \xi'^o_e+\tilde{\varphi}_e)} \\\nonumber
&&\cdot \exp(-(\frac{\eta_e}{2t}-\frac{D-1}{2})\widetilde{\Theta}_e) \sum_{[\tilde{k}_e]}(\widetilde{\text{Poly}}(\tilde{k}_e))^2\left(\exp(-2t\tilde{k}_e^2)e^{\mathbf{i}\tilde{k}_e(\xi^o_e-\xi'^o_e +\tilde{\varphi}_e)}\right)\\\nonumber
&=&\frac{\sqrt{\pi}}{\sqrt{2 t}}e^{\frac{(\eta_e)^2+(\eta'_e)^2+2t^2(D-1)^2}{4t}}e^{-t(\frac{\eta'_e}{2t}-\frac{\eta_e}{2t})^2 +2t(\frac{\eta'_e}{4t}-\frac{\eta_e}{4t}-\frac{\widetilde{\Theta}_e}{4t})^2} e^{\mathbf{i}(\frac{\eta_e}{4t}-\frac{D-1}{2}+\frac{\eta'_e}{4t}-\frac{\widetilde{\Theta}_e}{4t}) (\xi^o_e-\xi'^o_e+\tilde{\varphi}_e)}e^{-(\frac{\eta_e}{2t}-\frac{D-1}{2})\widetilde{\Theta}_e} \\\nonumber
&&\cdot\sum_{n=-\infty}^{\infty} \check{\text{Poly}}(2\pi n-(\xi^o_e-\xi'^o_e+\tilde{\varphi}_e))\exp(-\frac{(2\pi n-(\xi^o_e-\xi'^o_e+\tilde{\varphi}_e))^2}{8t})e^{\mathbf{i}2\pi n\text{mod}(\tilde{k}_e,1)}\\\nonumber
&\stackrel{t\rightarrow 0}{=}&\frac{\sqrt{\pi}}{\sqrt{2 t}}e^{\frac{(\eta_e)^2+(\eta'_e)^2+2t^2(D-1)^2}{4t}}e^{-t(\frac{\eta'_e}{2t}-\frac{\eta_e}{2t})^2 +2t(\frac{\eta'_e}{4t}-\frac{\eta_e}{4t}-\frac{\widetilde{\Theta}_e}{4t})^2} e^{\mathbf{i}(\frac{\eta_e}{4t}-\frac{D-1}{2}+\frac{\eta'_e}{4t}-\frac{\widetilde{\Theta}_e}{4t}) (\xi^o_e-\xi'^o_e+\tilde{\varphi}_e)} \\\nonumber
&&\cdot e^{-(\frac{\eta_e}{2t}-\frac{D-1}{2})\widetilde{\Theta}_e}\exp(-\frac{(\xi^o_e-\xi'^o_e+\tilde{\varphi}_e)^2}{8t})f_{\text{Poly}}(\frac{\eta'_e}{t}, \frac{\eta_e}{t},\frac{\widetilde{\Theta}_e}{t})
\end{eqnarray}
for large $\eta_e, \eta'_e$, where $\tilde{\varphi}_e:=\varphi(u_e,u'_e) +\varphi(\tilde{u}_e,\tilde{u}'_e)$, 
$\tilde{k}_e:=d_{N_e}- \frac{\eta'_e}{4t}-\frac{\eta_e}{4t}+\frac{\widetilde{\Theta}_e}{4t}=[\tilde{k}_e]+\text{mod}(\tilde{k}_e,1)$ with $[\tilde{k}_e]$ being the maximum integer less than or equal to $\tilde{k}_e$ and $\text{mod}(\tilde{k}_e,1)$ being the corresponding remainder, $\widetilde{\text{Poly}}(\tilde{k}_e)=\dim(\pi_{N_e})$ is a polynomial of $\tilde{k}_e$, and $f_{\text{Poly}}(\frac{\eta'_e}{t}, \frac{\eta_e}{t},\frac{\widetilde{\Theta}_e}{t})$ is a polynomial of the three variables $\frac{\eta'_e}{t}, \frac{\eta_e}{t},\frac{\widetilde{\Theta}_e}{t}$ satisfying
\begin{equation}
f_{\text{Poly}}(\frac{\eta'_e}{t}, \frac{\eta_e}{t},\frac{\widetilde{\Theta}_e}{t})\stackrel{t\rightarrow 0}{=}(\breve{\text{Poly}}(\frac{\eta'_e}{4t}+\frac{\eta_e}{4t}-\frac{\widetilde{\Theta}_e}{4t}))^2
\end{equation}
for large $\eta_e, \eta'_e$ and $\widetilde{\Theta}_e\ll \eta_e+\eta'_e$. Especially, we used the Poisson summation formula to get the third ``='' in Eq.\eqref{overinner}, which reads
\begin{equation}
  \sum_{n=-\infty}^{\infty}f(n)=2\pi\sum_{n=-\infty}^\infty\tilde{f}(2\pi n), \quad \tilde{f}(k):=\int_{\mathbb{R}}\frac{dx}{2\pi}e^{-\mathbf{i}kx}f(x), \quad n\in\mathbb{Z}
\end{equation}
with
\begin{equation}
f(x)=(\widetilde{\text{Poly}}(x+\text{mod}(\tilde{k}_e,1)))^2\left(\exp(-2t(x+\text{mod}(\tilde{k}_e,1))^2)e^{\mathbf{i} (x+\text{mod}(\tilde{k}_e,1)) (\xi^o_e-\xi'^o_e+\tilde{\varphi}_e)}\right),\  \   x=[\tilde{k}_e]
\end{equation}
and
\begin{equation}
\tilde{f}(k)=\frac{1}{2\sqrt{2\pi t}} \check{\text{Poly}}(k-(\xi^o_e-\xi'^o_e+\tilde{\varphi}_e))\exp(-\frac{(k-(\xi^o_e-\xi'^o_e+\tilde{\varphi}_e))^2}{8t})e^{\mathbf{i}k\text{mod}(\tilde{k}_e,1)},
\end{equation}
where $[\tilde{k}_e]$ takes the range of all integers for large $\eta_e, \eta'_e$ and $t\rightarrow0$, and $ \check{\text{Poly}}(x)$ is a polynomial given by
\begin{equation}
\check{\text{Poly}}(x)=\frac{1}{2\sqrt{2\pi t}}\left((\mathbf{i})^n a_n\frac{d^n}{dx^n} \exp(-\frac{x^2}{8t})+(\mathbf{i})^{n-1} a_{n-1}\frac{d^{n-1}}{dx^{n-1}} \exp(-\frac{x^2}{8t})+...+a_0 \exp(-\frac{x^2}{8t})\right) \exp(\frac{x^2}{8t})
\end{equation}
with $\widetilde{\text{Poly}}(x)$ expanded as $\widetilde{\text{Poly}}(x)=a_nx^n+a_{n-1}x^{n-1}+...+a_0$. Then, the overlap function is expressed as
\begin{eqnarray}\label{overlap2}
&&i^t(\mathbb{H}^o_e, \mathbb{H}'^o_e):=\frac{|\langle \breve{\Psi}_{\mathbb{H}^o_e},\breve{\Psi}_{\mathbb{H}'^o_e}\rangle|^2}{||\breve{\Psi}_{\mathbb{H}^o_e}||^2 ||\breve{\Psi}_{\mathbb{H}'^o_e}||^2}\\\nonumber
&\stackrel{t\rightarrow 0}{=}&\frac{ (f_{\text{Poly}}(\frac{\eta'_e}{t}, \frac{\eta_e}{t},\frac{\widetilde{\Theta}_e}{t}))^2}{(\breve{\text{Poly}}(\frac{\eta_e}{2t}))^2 (\breve{\text{Poly}}(\frac{\eta'_{e}}{2t}))^2} e^{-2t(\frac{\eta'_e}{2t}-\frac{\eta_e}{2t})^2 +4t(\frac{\eta'_e}{4t}-\frac{\eta_e}{4t}-\frac{\widetilde{\Theta}_e}{4t})^2} e^{-2(\frac{\eta_e}{2t}-\frac{D-1}{2})\widetilde{\Theta}_e}\exp(-\frac{(\xi^o_e-\xi'^o_e+\tilde{\varphi}_e)^2}{4t})
\end{eqnarray}
for large $\eta_e, \eta'_e$ and $\widetilde{\Theta}_e\ll \eta_e+\eta'_e$.
Now we can analyze the peakedness property of the overlap function for the case with $\widetilde{\Theta}_e\ll \eta_e+\eta'_e$. For the overlap function $i^t(\mathbb{H}^o_e, \mathbb{H}'^o_e)$ given by Eq.\eqref{overlap2}, we first conclude that it is sharply peaked at $\widetilde{\Theta}_e=0$ because of the factor $e^{-2(\frac{\eta_e}{2t}-\frac{D-1}{2})\widetilde{\Theta}_e}$. Moreover, noticing that $\tilde{\varphi}_e=0$ if $\widetilde{\Theta}_e=0$ by their definitions, one can further get that the overlap function $i^t(\mathbb{H}^o_e, \mathbb{H}'^o_e)$ is sharply peaked at $\xi^o_e=\xi'^o_e$ and $\eta_e=\eta'_e$ based on the factors $\exp(-\frac{(\xi^o_e-\xi'^o_e+\tilde{\varphi}_e)^2}{4t})$ and $e^{-2t(\frac{\eta'_e}{2t}-\frac{\eta_e}{2t})^2 +4t(\frac{\eta'_e}{4t}-\frac{\eta_e}{4t}-\frac{\widetilde{\Theta}_e}{4t})^2}$ respectively.

Let us consider the case with $\widetilde{\Theta}_e\simeq\eta_e+\eta'_e$ or $\widetilde{\Theta}_e\gg \eta_e+\eta'_e$. In this case, we have
 \begin{eqnarray}
&&\langle \breve{\Psi}_{\mathbb{H}^o_e},\breve{\Psi}_{\mathbb{H}'^o_e}\rangle\\\nonumber
&\stackrel{t\rightarrow 0}{=}&e^{\frac{(\eta_e)^2+(\eta'_e)^2+2t^2(D-1)^2}{4t}}\sum_{N_e}(\dim(\pi_{N_e}))^2\\\nonumber
&&\cdot\exp(-t(\frac{\eta_e}{2t}-d_{N_e})^2 -t(\frac{\eta'_e}{2t}-d_{N_e})^2)e^{\mathbf{i}N_e(\xi^o_e-\xi'^o_e+\varphi(u_e,u'_e) +\varphi(\tilde{u}_e,\tilde{u}'_e))}\exp(- N_e\widetilde{\Theta}_e)\\\nonumber
&\stackrel{t\rightarrow 0}{<}&e^{\frac{(\eta_e)^2+(\eta'_e)^2+2t^2(D-1)^2}{4t}}\exp(-t(\frac{\eta_e}{2t}-\frac{D+1}{2})^2 -t(\frac{\eta'_e}{2t}-\frac{D+1}{2})^2)\\\nonumber
&&+e^{\frac{(\eta_e)^2+(\eta'_e)^2+2t^2(D-1)^2}{4t}}[\eta_e/4t] \exp(-t(\frac{\eta_e}{4t}-\frac{D+1}{2})^2 -t(\frac{\eta'_e}{2t}-\frac{\eta_e}{4t}-\frac{D+1}{2})^2)(\breve{\text{Poly}}(\frac{\eta_e}{4t}))^2\exp(- \widetilde{\Theta}_e)\\\nonumber
&&+e^{\frac{(\eta_e)^2+(\eta'_e)^2+2t^2(D-1)^2}{4t}}\sum_{N_e=[\frac{\eta_e}{4t}]+1}^{+\infty} (\dim(\pi_{N_e}))^2\left(\exp(-t(\frac{\eta_e}{2t}-d_{N_e})^2 -t(\frac{\eta'_e}{2t}-d_{N_e})^2) \exp(- [\frac{\eta_e}{4t}]\widetilde{\Theta}_e)\right)
\\\nonumber
&\stackrel{t\rightarrow 0}{\simeq}&e^{\frac{(\eta_e)^2+(\eta'_e)^2+2t^2(D-1)^2}{4t}}\exp(-t(\frac{\eta_e}{2t}-\frac{D+1}{2})^2 -t(\frac{\eta'_e}{2t}-\frac{D+1}{2})^2)\\\nonumber
&&+e^{\frac{(\eta_e)^2+(\eta'_e)^2+2t^2(D-1)^2}{4t}}[\eta_e/4t] \exp(-t(\frac{\eta_e}{4t}-\frac{D+1}{2})^2 -t(\frac{\eta'_e}{2t}-\frac{\eta_e}{4t}-\frac{D+1}{2})^2)(\breve{\text{Poly}}(\frac{\eta_e}{4t}))^2\exp(- \widetilde{\Theta}_e)\\\nonumber
&&+\sqrt{\frac{\pi}{2t}}e^{\frac{(\eta_e)^2+(\eta'_e)^2+2t^2(D-1)^2}{4t}} (\breve{\text{Poly}}(\frac{\eta_e}{4t}+\frac{\eta'_e}{4t}))^2e^{-\frac{t}{2}(\frac{\eta'_e}{2t}-\frac{\eta_e}{2t})^2}  \exp(- [\frac{\eta_e}{4t}]\widetilde{\Theta}_e)\\\nonumber
\end{eqnarray}
for large $\eta_e$ and $\eta'_e$, where we analyzed the summation over $N_e$ in the ranges $\{0\}$, $\{1,...,[\frac{\eta_e}{4t}]\}$ and $\{[\frac{\eta_e}{4t}]+1,...,+\infty\}$ of $N_e$ separately. Again, we used the Poisson summation formula in the third step. Then, the Overlap Function is expressed as
\begin{eqnarray}\label{overlap1}
&&i^t(\mathbb{H}^o_e, \mathbb{H}'^o_e):=\frac{|\langle \breve{\Psi}_{\mathbb{H}^o_e},\breve{\Psi}_{\mathbb{H}'^o_e}\rangle|^2}{||\breve{\Psi}_{\mathbb{H}^o_e}||^2 ||\breve{\Psi}_{\mathbb{H}'^o_e}||^2}\\\nonumber
&\stackrel{t\rightarrow 0}{\lesssim}&\frac{\left(\sqrt{\frac{2t}{\pi}}\left(e^{-t((\frac{\eta_e}{2t})^2+ (\frac{\eta'_e}{2t})^2)}+f(\eta_e,\eta'_e)e^{-t(\frac{\eta_e}{4t})^2}e^{- \widetilde{\Theta}_e}\right) + (\breve{\text{Poly}}(\frac{\eta_e}{4t}+\frac{\eta'_e}{4t}))^2e^{-\frac{t}{2}(\frac{\eta'_e}{2t}-\frac{\eta_e}{2t})^2} e^{- [\frac{\eta_e}{4t}]\widetilde{\Theta}_e}\right)^2}{ (\breve{\text{Poly}}(\frac{\eta_e}{2t}))^2 (\breve{\text{Poly}}(\frac{\eta'_e}{2t}))^2}
\end{eqnarray}
for large $\eta_e$ and $\eta'_e$, with $f(\eta_e,\eta'_e)=[\eta_e/4t] \exp( -t(\frac{\eta'_e}{2t}-\frac{\eta_e}{4t}-\frac{D+1}{2})^2)(\breve{\text{Poly}}(\frac{\eta_e}{4t}))^2$.
Note that we considered $\widetilde{\Theta}_e\simeq\eta_e+\eta'_e$ or $\widetilde{\Theta}_e\gg \eta_e+\eta'_e$ here, so it is easy to see that the overlap function $i^t(\mathbb{H}^o_e, \mathbb{H}'^o_e)$ is suppressed exponentially by the factors $e^{-t((\frac{\eta_e}{2t})^2+ (\frac{\eta'_e}{2t})^2)}$, $e^{-t(\frac{\eta_e}{4t})^2}$ and $e^{- [\frac{\eta_e}{4t}]\widetilde{\Theta}_e}$  in Eq.\eqref{overlap1}.

Finally, by combining the analysis of the overlap function $i^t(\mathbb{H}^o_e, \mathbb{H}'^o_e)$ given by Eqs.\eqref{overlap2} and \eqref{overlap1}, one can gets that the overlap function $i^t(\mathbb{H}^o_e, \mathbb{H}'^o_e)$ is sharply peaked at $\xi^o_e=\xi'^o_e$, $\eta_e=\eta'_e$ and $V_e=V'_e$, $\tilde{V}_e=\tilde{V}'_e$, which is referred to as the peakedness property.
\subsubsection{Peakedness in holonomy representation}
 We are interested in the limit $t\rightarrow 0$ of the probability distribution
 \begin{equation}
 p_{\mathbb{H}_e^o}^t(h_e):=\frac{| \breve{\Psi}_{\mathbb{H}^o_e}(h_e)|^2}{||\breve{\Psi}_{\mathbb{H}^o_e}||^2 }.
 \end{equation}
 We will show that it is peaked on a set of elements of $SO(D+1)$ given by $\{h_e=u_ee^{\xi^o\tau_o}\bar{u}_e\tilde{u}_e^{-1}|\bar{u}_e\in SO(D-1)_{\tau_o}\}$, where $ SO(D-1)_{\tau_o}$ is the maximal subgroup of $SO(D+1)$ fixing both $\delta_1^I$ and $\delta_2^I$. To achieve this result, we take the expression
 \begin{eqnarray}\label{holopeak1}
&&\breve{\Psi}_{{\mathbb{H}}^o_e}({h}_e)\\\nonumber
&:=&\sum_{N_e}(\dim(\pi_{N_e}))^{3/2} e^{\frac{(\eta_e)^2+t^2(D-1)^2}{4t}}\left(\exp(-t(\frac{\eta_e}{2t}-d_{N_e})^2)e^{-\mathbf{i}\xi^o_ed_{N_e}}\langle N_e,\mathbf{N}|u_e^{-1}h_e\tilde{u}_e|N_e,\mathbf{N}\rangle\right.\\\nonumber
&&+\left.\exp(-t(\frac{\eta_e}{2t}+d_{N_e})^2)e^{\mathbf{i}\xi^o_ed_{N_e}}\langle N_e,\bar{\mathbf{N}}|u_e^{-1}h_e\tilde{u}_e|N_e,\bar{\mathbf{N}}\rangle\right)\\\nonumber
&\xlongequal[t\rightarrow 0 ]{\text{ large}\ \! \eta_e}&e^{\frac{(\eta_e)^2+t^2(D-1)^2}{4t}}e^{-\mathbf{i}\frac{(D-1)}{2}\xi^o_e}\sum_{N_e}(\dim(\pi_{N_e}))^{\frac{3}{2}} \exp(-t(\frac{\eta_e}{2t}-d_{N_e})^2)e^{-N_e \Theta_{u_e\!,\tilde{u}_e}(h_e)}e^{\mathbf{i}N_e(\varphi_{u_e\!,\tilde{u}_e}(h_e)-\xi^o_e)},
\end{eqnarray}
where $\Theta_{u_e\!,\tilde{u}_e}(h_e):=-\frac{\ln|\langle N_e,\mathbf{N}|u_e^{-1}h_e\tilde{u}_e|N_e,\mathbf{N}\rangle|}{N_e}$ and $e^{\mathbf{i}N_e\varphi_{u_e\!,\tilde{u}_e}(h_e)}:=\frac{\langle N_e,\mathbf{N}|u_e^{-1}h_e\tilde{u}_e|N_e,\mathbf{N}\rangle}{|\langle N_e,\mathbf{N}|u_e^{-1}h_e\tilde{u}_e|N_e,\mathbf{N}\rangle|}$. We need to consider the cases of $\Theta_{u_e\!,\tilde{u}_e}(h_e)\ll \eta_e$ and $\Theta_{u_e\!,\tilde{u}_e}(h_e)\simeq \eta_e$ or $\Theta_{u_e\!,\tilde{u}_e}(h_e)\gg \eta_e$ separately.

Let us first consider the case with $\Theta_{u_e,\tilde{u}_e}(h_e)\simeq \eta_e$ or $\Theta_{u_e,\tilde{u}_e}(h_e)\gg \eta_e$. In this case, we notice the factor $e^{-N_e \Theta_{u_e,\tilde{u}_e}(h_e)}$ in the final expression of \eqref{holopeak1}. Thus it is easy to see that this factor $e^{-N_e \Theta_{u_e,\tilde{u}_e}(h_e)}$ is sharply damped as $e^{-N_e \Theta_{u_e,\tilde{u}_e}(h_e)}\leq e^{-\frac{\eta_e}{4t}\Theta_{u_e,\tilde{u}_e}(h_e)}$ when $N_e\geq\frac{\eta_e}{4t}$ for large $\eta_e$ and $\Theta_{u_e,\tilde{u}_e}(h_e)\simeq \eta_e$ or $\Theta_{u_e,\tilde{u}_e}(h_e)\gg \eta_e$. Hence we have
 \begin{eqnarray}
&&\breve{\Psi}_{{\mathbb{H}}^o_e}({h}_e)\\\nonumber
&\stackrel{t\rightarrow 0}{\leq}&e^{\frac{(\eta_e)^2+t^2(D-1)^2}{4t}}e^{-\mathbf{i}\frac{(D-1)}{2}\xi^o_e}\sum^{+\infty}_{N_e=[\frac{\eta_e}{4t}]+1}(\dim(\pi_{N_e}))^{3/2} \exp(-t(\frac{\eta_e}{2t}-d_{N_e})^2)e^{-\frac{\eta_e}{4t} \Theta_{u_e,\tilde{u}_e}(h_e)}\\\nonumber
&&+e^{\frac{(\eta_e)^2+t^2(D-1)^2}{4t}}e^{-\mathbf{i}\frac{(D-1)}{2}\xi^o_e}\sum^{[\frac{\eta_e}{4t}]}_{N_e=0} (\dim(\pi_{N_e}))^{3/2} \exp(-t(\frac{\eta_e}{4t}-\frac{D+1}{2})^2)\\\nonumber
&\stackrel{t\rightarrow 0}{\lesssim}&\sqrt{\frac{\pi}{t}}e^{\frac{(\eta_e)^2+t^2(D-1)^2}{4t}}e^{-\mathbf{i}\frac{(D-1)}{2}\xi^o_e}e^{-\frac{\eta_e}{4t} \Theta_{u_e,\tilde{u}_e}(h_e)}(\breve{\text{Poly}}(\frac{\eta_e}{2t}))^{3/2}\\\nonumber
&&+([\frac{\eta_e}{4t}]+1)e^{\frac{(\eta_e)^2+t^2(D-1)^2}{4t}}e^{-\mathbf{i}\frac{(D-1)}{2}\xi^o_e}(\breve{\text{Poly}} (\frac{\eta_e}{4t}))^{3/2} \exp(-t(\frac{\eta_e}{4t}-\frac{D+1}{2})^2)
\end{eqnarray}
for large $\eta_e$, where we used
 \begin{equation}
\exp(-t(\frac{\eta_e}{2t}-d_{N_e})^2)\leq \exp(-t(\frac{\eta_e}{4t}-\frac{D+1}{2})^2),\quad (\dim(\pi_{N_e}))^{3/2} \leq(\breve{\text{Poly}}(\frac{\eta_e}{4t}))^{3/2}
 \end{equation}
and $e^{-N_e \Theta_{u_e\!,\tilde{u}_e}(h_e)}\leq1$ for $N_e\in\{0,...,[\frac{\eta_e}{4t}]\}$. Then the probability distribution can be evaluated by
\begin{eqnarray}
 &&p_{\mathbb{H}_e^o}^t(h_e):=\frac{| \breve{\Psi}_{\mathbb{H}^o_e}(h_e)|^2}{||\breve{\Psi}_{\mathbb{H}^o_e}||^2 }\\\nonumber
 &&\stackrel{t\rightarrow 0}{\lesssim}\sqrt{\frac{2t}{\pi}}(\sqrt{\frac{\pi}{t}}e^{-\frac{\eta_e}{4t} \Theta_{u_e,\tilde{u}_e}(h_e)}
+([\frac{\eta_e}{4t}]+1)\exp(-t(\frac{\eta_e}{4t}-\frac{D+1}{2})^2))^2\breve{\text{Poly}}(\frac{\eta_e}{2t}).
 \end{eqnarray}
By this expression, it is  easy to see that $p_{\mathbb{H}_e^o}^t(h_e)$ vanishes exponentially in the limit $t\rightarrow0$ for large $\eta_e$ and $\Theta_{u_e,\tilde{u}_e}(h_e)\simeq \eta_e$ or $\Theta_{u_e,\tilde{u}_e}(h_e)\gg \eta_e$.

 Now let us turn to the case with $\Theta_{u_e,\tilde{u}_e}(h_e)\ll \eta_e$. In this case, we have
\begin{eqnarray}
&&\breve{\Psi}_{{\mathbb{H}}^o_e}({h}_e)\\\nonumber
&\xlongequal[t\rightarrow 0 ]{\text{ large}\ \! \eta_e}&e^{\frac{(\eta_e)^2+t^2(D-1)^2}{4t}}e^{-\mathbf{i}\frac{(D-1)}{2}\xi^o_e}\sum_{N_e}(\dim(\pi_{N_e}))^{\frac{3}{2}} \exp(-t(\frac{\eta_e}{2t}-d_{N_e})^2)e^{-N_e \Theta_{u_e,\tilde{u}_e}(h_e)}e^{\mathbf{i}N_e(\varphi_{u_e,\tilde{u}_e}(h_e)-\xi^o_e)}\\\nonumber
&{=}&e^{\frac{(\eta_e)^2+t^2(D-1)^2}{4t}} e^{-\mathbf{i}\frac{(D-1)}{2}\xi^o_e} \exp(-t(\frac{\eta_e}{2t}-\frac{D-1}{2})^2+t(\frac{\eta_e}{2t}-\frac{D-1}{2}-\frac{\Theta_{u_e,\tilde{u}_e}(h_e)}{2t})^2)\\\nonumber
&&\cdot e^{\mathbf{i}(\frac{\eta_e}{2t}-\frac{D-1}{2}-\frac{\Theta_{u_e,\tilde{u}_e}(h_e)}{2t})(\varphi_{u_e,\tilde{u}_e}(h_e)-\xi^o_e)}\sum_{N_e}(\dim(\pi_{N_e}))^{3/2} \exp(-t\breve{k}_e^2)e^{\mathbf{i}\breve{k}_e(\varphi_{u_e,\tilde{u}_e}(h_e)-\xi^o_e)}\\\nonumber
&\xlongequal[t\rightarrow 0 ]{\text{ large}\ \! \eta_e}&\sqrt{\frac{\pi}{t}}e^{\frac{(\eta_e)^2+t^2(D-1)^2}{4t}} e^{-\mathbf{i}\frac{(D-1)}{2}\xi^o_e} \exp(-t(\frac{\eta_e}{2t}-\frac{D-1}{2})^2+t(\frac{\eta_e}{2t}-\frac{D-1}{2}-\frac{\Theta_{u_e,\tilde{u}_e}(h_e)}{2t})^2)\\\nonumber
&&\cdot \ e^{\mathbf{i}(\frac{\eta_e}{2t}-\frac{D-1}{2}-\frac{\Theta_{u_e,\tilde{u}_e}(h_e)}{2t})(\varphi_{u_e,\tilde{u}_e}(h_e)-\xi^o_e)} (\breve{\text{Poly}}(\frac{\eta_e}{2t}))^{3/2}\\\nonumber
&&\cdot\sum_{n=-\infty}^{\infty} \exp(-\frac{(2\pi n-(\varphi_{u_e,\tilde{u}_e}(h_e)-\xi^o_e))^2}{4t})e^{\mathbf{i}2\pi n(\text{mod}(\breve{k}_e,1))}\\\nonumber
&\stackrel{t\rightarrow 0}{=}&\sqrt{\frac{\pi}{t}}e^{\frac{(\eta_e)^2+t^2(D-1)^2}{4t}} e^{-\mathbf{i}\frac{(D-1)}{2}\xi^o_e} \exp(-t(\frac{\eta_e}{2t}-\frac{D-1}{2})^2+t(\frac{\eta_e}{2t}-\frac{D-1}{2}-\frac{\Theta_{u_e,\tilde{u}_e}(h_e)}{2t})^2)\\\nonumber
&&\cdot e^{\mathbf{i}(\frac{\eta_e}{2t}-\frac{D-1}{2}-\frac{\Theta_{u_e,\tilde{u}_e}(h_e)}{2t})(\varphi_{u_e,\tilde{u}_e}(h_e)-\xi^o_e)} (\breve{\text{Poly}}(\frac{\eta_e}{2t}))^{3/2} \exp(-\frac{(\varphi_{u_e,\tilde{u}_e}(h_e)-\xi^o_e)^2}{4t}),
\end{eqnarray}
where $\breve{k}_e=[\breve{k}_e]+\!\mod(\breve{k}_e,1):= -\frac{\eta_e}{2t}+d_{N_e}+\frac{\Theta_{u_e,\tilde{u}_e}(h_e)}{2t}$ and $ \breve{\text{Poly}}(N_e)=\dim(\pi_{N_e})$. Then, we have
\begin{eqnarray}\label{ph}
&& p_{\mathbb{H}_e^o}^t(h_e):=\frac{| \breve{\Psi}_{\mathbb{H}^o_e}(h_e)|^2}{||\breve{\Psi}_{\mathbb{H}^o_e}||^2 } \\\nonumber
   && \xlongequal[t\rightarrow 0 ]{\text{ large}\ \! \eta_e} \sqrt{2\pi/t} \exp\left((-\frac{\eta_e}{t}+{D-1}+\frac{\Theta_{u_e,\tilde{u}_e}(h_e)}{2t})\Theta_{u_e,\tilde{u}_e}(h_e)\right)
 \breve{\text{Poly}}(\frac{\eta_e}{2t}) \exp(-\frac{(\varphi_{u_e,\tilde{u}_e}(h_e)-\xi^o_e)^2}{2t})
\end{eqnarray}
for $\Theta_{u_e,\tilde{u}_e}(h_e)\ll \eta_e$.
Let us analyze the peakedness property of $p_{\mathbb{H}_e^o}^t(h_e)$ in the case with $\eta_e$ being large and $\Theta_{u_e,\tilde{u}_e}(h_e)\ll \eta_e$. According to Eq.\eqref{ph}, $p_{\mathbb{H}_e^o}^t(h_e)$ is sharply peaked at $\Theta_{u_e,\tilde{u}_e}(h_e)=0$ and $\varphi_{u_e,\tilde{u}_e}(h_e)=\xi^o_e$ in the limit $t\rightarrow0$ by the factor $ \exp\left((-\frac{\eta_e}{t}+{D-1}+\frac{\Theta_{u_e,\tilde{u}_e}(h_e)}{2t})\Theta_{u_e,\tilde{u}_e}(h_e)\right)$ and $\exp(-\frac{(\varphi_{u_e,\tilde{u}_e}(h_e)-\xi^o_e)^2}{2t})$ respectively.

Finally, let us combine the results in the two cases discussed above.
Then one can conclude that the probability distribution $p_{\mathbb{H}_e^o}^t(h_e)$ is sharply peaked at $\Theta_{u_e,\tilde{u}_e}(h_e)=0$ and $\varphi_{u_e,\tilde{u}_e}(h_e)=\xi^o_e$, which corresponds to that the holonomy $h_e$ takes values in a set of elements of $SO(D+1)$ given by $\{h_e=u_ee^{\xi^o\tau_o}\bar{u}_e\tilde{u}_e^{-1}|\bar{u}_e\in SO(D-1)_{\tau_o}\}$, with $ SO(D-1)_{\tau_o}$ being the maximal subgroup of $SO(D+1)$ fixing both $\delta_1^I$ and $\delta_2^I$.
\subsubsection{Peakedness in momentum representation}
With the edge simplicity constraint being solved, the basis of the momentum representation space can be given by $\{|N_e, V_e,\tilde{V}_e \rangle\}$, where $|N_e, V_e,\tilde{V}_e \rangle$ corresponds to the matrix-element function $\langle N_e, V_e|h_e|N_e,-\tilde{V}_e\rangle$ of $h_e$, selected by the Perelomov type coherent states of $SO(D+1)$. Then we define the momentum representation of a state $\varphi$ by
\begin{equation}
\varphi(N_e, V_e,\tilde{V}_e):=\langle N_e, V_e,\tilde{V}_e|\varphi\rangle.
\end{equation}
It is easy to calculate
\begin{eqnarray}
&&\breve{\Psi}_{{\mathbb{H}}^o_e}(N_e, V'_e,\tilde{V}'_e)=\int_{SO(D+1)}dh_e\breve{\Psi}_{{\mathbb{H}}^o_e}({h}_e) \overline{\langle N_e, V'_e|h_e|N_e,-\tilde{V}'_e\rangle}\\\nonumber
&\xlongequal[t\rightarrow 0 ]{\text{ large}\ \! \eta_e}&(\dim(\pi_{N_e}))^{1/2} e^{\frac{(\eta_e)^2+t^2(D-1)^2}{4t}}e^{-\mathbf{i}\frac{(D-1)}{2}\xi^o_e}\exp(-t(\frac{\eta_e}{2t}-d_{N_e})^2-N_e(\Theta(V_e,V'_e) +\Theta(\tilde{V}_e,\tilde{V}'_e)))\\\nonumber
&&\cdot e^{\mathbf{i}N_e(-\xi^o_e+\varphi(\tilde{V}_e,\tilde{V}'_e)+\varphi(V_e,{V'}_e))}
\end{eqnarray}
where we define $\Theta(V_e,V'_e):=-\frac{\ln|\langle N_e, V_e|N_e, V'_e\rangle|}{N_e}\geq0$, $\Theta(\tilde{V}_e,\tilde{V}'_e):=-\frac{\ln|\langle N_e, -\tilde{V}'_e|N_e,-\tilde{ V}_e\rangle|}{N_e}\geq0$ and  $e^{\mathbf{i}N_e\varphi(V_e,{V'}_e)}:=\frac{\langle N_e, V_e|N_e, V'_e\rangle}{|\langle N_e, V_e|N_e, V'_e\rangle|}$, $e^{\mathbf{i}N_e\varphi(\tilde{V}_e,\tilde{V}'_e)}:=\frac{\langle N_e, -\tilde{V}'_e|N_e, -\tilde{V}_e\rangle}{|\langle N_e, -\tilde{V}'_e|N_e, -\tilde{V}_e\rangle|}$. These variables satisfy the relations that $\Theta(V_e,V'_e)=0$, $\Theta(\tilde{V}_e,\tilde{V}'_e)=0$ if and only if $V_e=V'_e$, $\tilde{V}_e=\tilde{ V}'_e$ respectively, and $\varphi(V_e,{V'}_e)=0$, $\varphi(\tilde{V}_e,\tilde{V}'_e)=0$ for $V_e=V'_e$, $\tilde{V}_e=\tilde{ V}'_e$ respectively.
 We are interested in the probability amplitude
\begin{eqnarray}\label{pth}
&&p^t_{{\mathbb{H}}^o_e}(N_e, V'_e,\tilde{V}'_e):=\frac{|\breve{\Psi}_{{\mathbb{H}}^o_e}(N_e, V'_e,\tilde{V}'_e)|^2}{||\breve{\Psi}_{{\mathbb{H}}^o_e}||^2}\\\nonumber
&\xlongequal[t\rightarrow 0 ]{\text{ large}\ \! \eta_e}&\frac{\dim(\pi_{N_e}) \exp(-2t(\frac{\eta_e}{2t}-d_{N_e})^2-2N_e(\Theta(V_e,V'_e) +\Theta(\tilde{V}_e,\tilde{V}'_e)))}{\sum_{N'_e}(\dim(\pi_{N'_e}))^2 \exp(-2t(\frac{\eta_e}{2t}-d_{N'_e})^2)}\\\nonumber
&\xlongequal[t\rightarrow 0 ]{\text{ large}\ \! \eta_e}&\frac{\dim(\pi_{N_e}) \exp(-2t(\frac{\eta_e}{2t}-d_{N_e})^2)\exp(-2N_e(\Theta(V_e,V'_e) +\Theta(\tilde{V}_e,\tilde{V}'_e))) }{(\breve{\text{Poly}}(\frac{\eta_e}{2t}))^2\sqrt{\frac{\pi}{2t}}}.
\end{eqnarray}
One should notice that in the limit $t\rightarrow0$, the factor $\frac{\exp(-2t(\frac{\eta_e}{2t}-d_{N_e})^2)}{(\breve{\text{Poly}}(\frac{\eta_e}{2t}))^2\sqrt{\frac{\pi}{2t}}}$ is peaked at $|\frac{\eta_e}{2t}-d_{N_e}|\ll\frac{1}{2t}$ and the shape of this peak sinks with $t\rightarrow0$. Moreover, the factor $\exp(-2N_e(\Theta(V_e,V'_e) +\Theta(\tilde{V}_e,\tilde{V}'_e)))$ is sharply peaked at $\Theta(V_e,V'_e) +\Theta(\tilde{V}_e,\tilde{V}'_e)=0$ for $|\frac{\eta_e}{2t}-d_{N_e}|\ll\frac{1}{2t}$, $t\rightarrow0$ and large $\eta_e$. Combing these results and according to Eq.\eqref{pth}, we get that the peak of probability amplitude $p^t_{{\mathbb{H}}^o_e}(N_e, V'_e,\tilde{V}'_e)$ has different behaviours for $N_e$ and $(\tilde{V}'_e,{V}'_e)$ in the limit $t\to 0$. For large $\eta_e$, this peak locates at $|\frac{\eta_e}{2t}-d_{N_e}|\ll\frac{1}{2t}$ and $V'_e =V_e$, $\tilde{V}'_e=\tilde{ V}_e$. Moreover, this peak sinks and moves to $N_e\to \infty$ with $t\to 0$ in the perspective of $N_e$, while this peak grows with $t\rightarrow0$ in the perspective of $(\tilde{V}'_e,{V}'_e)$.

\section{Conclusion and Discussion}\label{sec5}
Based on the generalized twisted geometric parametrization of the $SO(D+1)$ holonomy-flux phase space, the twisted geometry coherent states in all dimensional LQG are constructed by composing the two families of Perelomov type coherent state of $SO(D+1)$ properly. It was shown that the twisted geometry coherent states provide a resolution of identity of the Hilbert space with edge simplicity constraint being solved for all dimensional LQG. Moreover, taking advantage of its Gaussian superposition formulation, we showed that the expectation values of holonomy and flux operator with respect to the twisted geometry coherent state coincide with the corresponding classical values given by the labels of the coherent states, up to some gauge degrees of freedom. Besides, the peakedness properties of the twisted geometry coherent state were studied, i.e. the wave functions of the twisted geometry coherent state have well-behaved peakedness properties in holonomy, momentum and phase space representations. The peakedness in various representations and the (relative) uncertainty of the expectation values are well controlled by the semi-classical parameter $t$. Therefore the twisted geometry coherent state comprises a candidate for the semi-classical analysis of all dimensional LQG.

There still remains some issues worth being discussed. First, the expectation value of holonomy operator is not exactly identical to its classical value given by the labels of the coherent state. As shown in our paper, it can be regarded that each twisted geometry coherent state is labeled by a family of classical holonomies distinguished with each other by the gauge components of the simplicity constraint, while the expectation value of the holonomy operator is independent of the undetermined gauge components in these classical holonomies.
This result would be helpful in the treatment of the regularization of the scalar constraint in all dimensional LQG.  Following the standard regularization methods in LQG, the classical scalar constraint can be given by the summation of the Euclidean term, the Lorentzian term and an additional unmanageable term proportioning to the gauge component with respect to simplicity constraint, which is used to offset the gauge (with respect to simplicity constraint) variant component in the holonomies appearing in the Euclidean term. Based on the result of the expectation value of holonomy operator, the quantum  scalar constraint operator could be simplified in the twisted geometry coherent state basis. In other words, since the expectation value of the holonomy operator for the twisted geometry coherent state is independent of the gauge (with respect to simplicity constraint) component, the unmanageable term used to offset the gauge variant component in the holonomies might be unnecessary. Second, the calculations and results based on the twisted geometry coherent state can be extended to the superposition type coherent state easily. At first, we note that the twisted geometry coherent states recover some of the superposition type coherent states up to some prefactors for $t\rightarrow0$ and large $\eta_e$, and our calculation are also proceeded for the case of $t\rightarrow0$ and large $\eta_e$. Moreover, the procedures of the calculations are dominated by the exponential factors $\exp(-2t(\frac{\eta_e}{2t}-d_{N_e}))$ of $N_e$. Thus, the results are independent of the prefactors at leading order of $t$. Third, one can expect further studies of the property of twisted geometry coherent state. An useful one is the ``Ehrenfest Property'', which would be helpful to evaluate the expectation value of more complicated operators, e.g. the scalar constraint operator. Finally, our study can be generalized to other compact Lie groups by adopting the relation $SU(N)\subset U(N)\subset SO(2N)$, which will be  useful in the corresponding quantum gauge field theories.
\section*{Acknowledgments}
This work is supported by the National Natural Science Foundation of China (NSFC) with Grants No. 12047519, No. 11775082, No. 11875006 and No. 11961131013. G.L. acknowledges the support by China Postdoctoral Science Foundation with Grant No. 2021M691072. C. Z. acknowledges the support by the Polish Narodowe Centrum Nauki, Grant No. 2018/30/Q/ST2/00811.
\appendix
\section{Poisson Summation Formula and associated calculations}\label{app1}
The Poisson summation formula is given as follows. Let $f$ be an $L_1(\mathbb{R}, dx)$ function such that the series
\begin{equation}
\phi(y)=\sum_{n=-\infty}^{\infty}f(y+ns)
\end{equation}
is absolutely and uniformly convergent for $y\in [0, s], s > 0$. Then
\begin{equation}
  \sum_{n=-\infty}^{\infty}f(ns)=\frac{2\pi}{s}\sum_{n=-\infty}^\infty\tilde{f}(\frac{2\pi n}{s}),
\end{equation}
where $  \tilde{f}(k):=\int_{\mathbf{R}}\frac{dx}{2\pi}e^{-\mathbf{i}kx}f(x)$ is the Fourier transform of $f$. The proof of this theorem can be found in \cite{Poissonsummation}. In this paper, the application of the Poisson summation formula is involved in the calculation of the expression
\begin{equation}
\sum_{N=0}^{\infty}F(N)\exp(-2t({\frac{\eta}{2t}-\frac{D-1}{2}}-N)^2).
\end{equation}
Let us consider three cases of $F(x)$ separately. \\
\textbf{Case I:} $F(x)=x^\ell$ is a polynomial with $\ell\in\mathbb{N}$. Define $\alpha_\eta:={\frac{\eta}{2t}-\frac{D-1}{2}}$, let us consider the following calculations.
\begin{eqnarray}\label{xell}
&&\sum_{N=0}^{\infty}F(N)\exp(-2t({N}-\alpha_\eta)^2)\\\nonumber
&=&\sum_{k={-\alpha_\eta}}^{\infty}(\alpha_\eta+k)^\ell\exp(-2tk^2)\\\nonumber
&=&(\alpha_\eta)^\ell\sum_{[k]=[-\alpha_\eta]}^{\infty}(1+\frac{([k]+r)}{\alpha_\eta})^\ell\exp(-2t([k]+r)^2),
\end{eqnarray}
where  $k:={N}-\alpha_\eta$, $r=-\alpha_\eta-[-\alpha_\eta]$ and $[k]$ represents the maximal integer no greater than $k$. Note that we have
\begin{eqnarray}
&&\alpha_\eta ^{-\ell'}\sum_{m=-\infty}^{\infty}(m+r)^{\ell'} e^{(-2t(m+r)^2)} =\alpha_\eta ^{-\ell'}\sum_{m=-\infty}^{\infty}e^{2\pi\mathbf{i}mr}\sqrt{\frac{\pi}{2t}}(\frac{\mathbf{i}}{2\pi})^{\ell'}\frac{d^{\ell'}} {dm^{\ell'}}e^{-\frac{( 2\pi m)^2}{8t}},\quad \ell'\in\mathbb{N}.
\end{eqnarray}
Then by setting $[k]=m$ in Eq.\eqref{xell}, we have
\begin{eqnarray}
\sum_{N=0}^{\infty}F(N)\exp(-2t(\alpha_\eta-N)^2)
\xlongequal[t\rightarrow 0 ]{\text{ large}\ \! \eta}\sqrt{\frac{\pi}{2t}}(\alpha_\eta)^\ell\left(1+\mathcal{O}(\frac{t}{\eta}) +\mathcal{O}(e^{-\frac{\pi^2}{2t}})\right).
\end{eqnarray}
\\ \textbf{Case II:} $F(x)=(P(x))^{1/4}$, where $P(x)$ is a polynomial of $x$ with $P(x)>0$ if $x>0$. Let us focus on the case of $\alpha_\eta>0, x>0$ involved in this paper. We can reformulate $F(x)$ as $F(x)=(P(\alpha_\eta))^{1/4}f(z)$ with $f(z):=(1+z)^{1/4}$ and $z:=\frac{P(x)}{P(\alpha_\eta)}-1>-1$. Then by Taylor's theorem, we have
\begin{equation}
f(z)=1+\sum_{n=1}^{\infty}\left(\begin{array}{lr}
q \\
 n
\end{array}
\right)z^n,\quad  \left(\begin{array}{lr}
q \\
 n
\end{array}
\right)=(-1)^{n+1}\frac{q(1-q)...(n-1+q)}{n!}
\end{equation}
with $q=1/4$ here.
To proceed next step of the calculation, we introduce a lemma as follows.
\\ \textbf{Lemma.} For each $l\geq0$ there exist $0<\beta_{l}<\infty$ such that
\begin{equation}
f_{2l+1}(z)-\beta_lz^{2l+2}\leq f(z)\leq f_{2l+1}(z),
\end{equation}
where $f_{l}(z)=1+\sum_{n=1}^{l}\left(\begin{array}{lr}
q \\
 n
\end{array}
\right)z^n,\quad  \left(\begin{array}{lr}
q \\
 n
\end{array}
\right)$ denotes the partial Taylor series of $f(z)=(1+z)^q$, $0<q\leq 1/4$, up to order $z^k$.

The proof of this lemma can be find in \cite{Giesel_2007}. Now let us set $x=N$ in $P(x)$ so that $z=\frac{P(N)}{P(\alpha_\eta)}-1$. Then by using the results of \textbf{Case I}, we can give
 \begin{eqnarray}
\sqrt{\frac{2t}{\pi}}\sum_{N=0}^{\infty}z^{\ell'}\exp(-2t(\alpha_\eta-N)^2)
\xlongequal[t\rightarrow 0 ]{\text{ large}\ \! \eta}\mathcal{O}(\frac{t}{\eta}) +\mathcal{O}(e^{-\frac{\pi^2}{2t}}).
\end{eqnarray}
Further by using the above \textbf{Lemma}, we get
\begin{eqnarray}
\sqrt{\frac{2t}{\pi}}\sum_{N=0}^{\infty}F(N)\exp(-2t(\alpha_\eta-N)^2)
\xlongequal[t\rightarrow 0 ]{\text{ large}\ \! \eta}(P(\alpha_\eta))^{1/4}(1+\mathcal{O}(\frac{t}{\eta}) +\mathcal{O}(e^{-\frac{\pi^2}{2t}})).
\end{eqnarray}
\\ \textbf{Case III:} $F(x)=\frac{x(2x+D-3)}{(x+D-2)(2x+D-1)}$ with $F(N)=|\alpha_1(N)|^2$. First, it is easy to see that
\begin{equation}
0<\sqrt{\frac{2t}{\pi}}\sum_{N=0}^{\infty}F(N)\exp(-2t(\alpha_\eta-N)^2)<\sqrt{\frac{2t}{\pi}}\sum_{N=0}^{\infty} \exp(-2t(\alpha_\eta-N)^2)\simeq1+\mathcal{O}(\frac{t}{\eta}) +\mathcal{O}(e^{-\frac{\pi^2}{2t}})
\end{equation}
for $t\rightarrow 0$ and $\eta $ being large.
Then, we can evaluate that
\begin{eqnarray}
 && \sqrt{\frac{2t}{\pi}}\sum_{N=0}^{\infty}(1-F(N))\exp(-2t(\alpha_\eta-N)^2)\\\nonumber
   &<& \sqrt{\frac{2t}{\pi}}\left(\sum_{N=0}^{[\frac{\eta}{4t}]}\exp(-2t(\alpha_\eta-N)^2)+ \sum_{N=[\frac{\eta}{4t}]+1}^{\infty}(1-F(\frac{\eta}{4t}))\exp(-2t(\alpha_\eta-N)^2)\right) \\
   &\stackrel{t\rightarrow 0}{\lesssim}&   \sqrt{\frac{2t}{\pi}}([\frac{\eta}{4t}]+1)\exp(-2t(\frac{\eta}{4t}-\frac{D-1}{2})^2)+ (1-F(\frac{\eta}{4t}))(1+\mathcal{O}(\frac{t}{\eta}) +\mathcal{O}(e^{-\frac{\pi^2}{2t}}))\\\nonumber
   &\stackrel{t\rightarrow 0}{\simeq}&\mathcal{O}(e^{-\frac{\eta^2}{8t}})+\mathcal{O}(\frac{t}{\eta})
\end{eqnarray}
for large $\eta$. Finally, for this case we can conclude that
\begin{equation}
\sqrt{\frac{2t}{\pi}}\sum_{N=0}^{\infty}F(N)\exp(-2t(\alpha_\eta-N)^2)\xlongequal[t\rightarrow 0 ]{\text{ large}\ \! \eta} 1+\mathcal{O}(e^{-\frac{\eta^2}{8t}})+\mathcal{O}(e^{-\frac{\pi^2}{2t}})+\mathcal{O}(\frac{t}{\eta}).
\end{equation}

\section{The diagonal matrix elements of holonomy operator in spin-network basis}\label{app2}
To give the diagonal matrix elements of holonomy operator in spin-network basis, let us first consider the Clebsh-Gordan coefficients related to the states in the simple representation space.  Recall the orthonormal basis $\{\Xi^{N,\mathbf{M}}(\bm{x})\}$ (or $\{|N,\mathbf{M}\rangle\}$ in Dirac bracket formulation) of the sphere harmonic function space on $S^D$. The Clebsh-Gordan coefficient can be given by
\begin{equation}
\langle N', \mathbf{M}';N'', \mathbf{M}''|N, \mathbf{M}\rangle \langle N, \mathbf{0}|N', \mathbf{0};N'', \mathbf{0}\rangle =\dim(\pi_{N})\int_{SO(D+1)}dg \overline{D_{(\mathbf{M},\mathbf{0})}^N(g)}D_{(\mathbf{M'},\mathbf{0})}^{N'} (g) D_{(\mathbf{M}'', \mathbf{0})}^{N''}(g),
\end{equation}
where $\dim(\pi_{N})=\frac{(D+N-2)!(2N+D-1)}{(D-1)!N!}$, $|N', \mathbf{M}';N'', \mathbf{M}''\rangle:=|N', \mathbf{M}'\rangle\otimes|N'', \mathbf{M}''\rangle$ and
\begin{equation}
D_{(\mathbf{M},\mathbf{0})}^N(g):=\langle N,\mathbf{M}|g|N, \mathbf{0}\rangle.
\end{equation}
Then we have
\begin{equation}
 |\langle N+1, \mathbf{0}|N, \mathbf{0};1, \mathbf{0}\rangle|^2 =\dim(\pi_{N+1})\int_{SO(D+1)}dg \overline{D_{(\mathbf{0},\mathbf{0})}^{N+1}(g)}D_{(\mathbf{0},\mathbf{0})}^{N} (g) D_{(\mathbf{0}, \mathbf{0})}^{1}(g).
\end{equation}
Let us note that \cite{vilenkin2013representation}
\begin{equation}
D_{(\mathbf{0},\mathbf{0})}^{N} (g) =D_{(\mathbf{0},\mathbf{0})}^{N} (\theta)=\frac{(D-2)!N!}{(D+N-2)!}C_N^{\frac{D-1}{2}}(\cos\theta) ,\quad C_1^{\frac{D-1}{2}}(\cos\theta)=(D-1)\cos\theta,
\end{equation}
and
\begin{equation}
C_{N+1}^{\frac{D-1}{2}}(\cos\theta) =\frac{2N+D-1}{N+1}\cos\theta C_{N}^{\frac{D-1}{2}}(\cos\theta) -\frac{N+D-2}{N+1} C_{N-1}^{\frac{D-1}{2}}(\cos\theta).
\end{equation}
Then we can calculate
\begin{equation}
 |\langle N+1, \mathbf{0}|N, \mathbf{0};1, \mathbf{0}\rangle|^2 =\frac{D+N-1}{2N+D-1},
\end{equation}
and similarly we have
\begin{equation}
 |\langle N-1, \mathbf{0}|N, \mathbf{0};1, \mathbf{0}\rangle|^2 =\frac{N}{2N+D-1},
\end{equation}
Notice the relation between the function $D_{(\mathbf{M},\mathbf{0})}^N(g)$ and $\Xi^{N,\mathbf{M}}(\bm{x})$, we have
\begin{eqnarray}\label{NNNNNN}
&&\langle N', \mathbf{M}';N'', \mathbf{M}''|N, \mathbf{M}\rangle \langle N, \mathbf{0}|N', \mathbf{0};N'', \mathbf{0}\rangle\\\nonumber& =&\sqrt{\frac{\dim(\pi_{N})}{\dim(\pi_{N'})\dim(\pi_{N''})}}
\int_{SO(D+1)} d\bm{x}\overline{\Xi^{N,\mathbf{M}}(\bm{x})} {\Xi^{N',\mathbf{M}''}(\bm{x}) }{\Xi^{N'',\mathbf{M}''}(\bm{x})}.
\end{eqnarray}

The normalized harmonic function $c_N(x_\imath+\mathbf{i}x_\jmath)$ can be denoted by $|N,V_{\imath\jmath}\rangle$ with $\imath, \jmath=1,...,D+1$, $(x_1,...,x_{D+1})\in S^D$, $V_{\imath\jmath}:=2\delta_\imath^{[I}\delta_\jmath^{J]}$ and $c_N$ being given by
\begin{equation}
c_N=\frac{2^N \Gamma(N+\frac{D-1}{2})}{\Gamma(\frac{D-1}{2})}\left(\frac{(D-2)!(2N+D-1)}{(2N+D-2)!(D-1)}\right)^{1/2}.
\end{equation}
A harmonic function basis of the definition representation space of $SO(D+1)$ is given as,
\begin{equation}
(x_1+\mathbf{i}x_2),\quad (x_1-\mathbf{i}x_2), \quad (x_3+\mathbf{i}x_4),..., (x_{D}+\mathbf{i}x_{D+1}),\quad (x_{D}-\mathbf{i}x_{D+1})
\end{equation}
for $D+1$ being even, and
\begin{equation}
(x_1+\mathbf{i}x_2),\quad (x_1-\mathbf{i}x_2), \quad (x_3+\mathbf{i}x_4),..., (x_{D-1}+\mathbf{i}x_{D}),\quad (x_{D-1}-\mathbf{i}x_{D}),\quad x_{D+1}
\end{equation}
for $D+1$ being odd.
 By using Dirac bracket notation, these functions can be expressed as following normalized states, which reads
\begin{equation}
\{|1, V_{\imath\jmath}\rangle\},\quad (\imath,\jmath)\in\{(1,2),(2,1),(3,4),(4,3),...,(D,D+1),(D+1,D)\}
\end{equation}
for $D+1$ being even, and
\begin{equation}
\{|1, V_{\imath\jmath}\rangle,\   |1,\delta_{D+1}\rangle\}, \quad (\imath,\jmath)\in\{(1,2),(2,1),(3,4),(4,3),...,(D-1,D),(D,D-1)\}
\end{equation}
for $D+1$ being odd.
It is easy to check that
\begin{equation}\label{CG00}
|1, V_{12};N,V_{12}\rangle:=|1, V_{12}\rangle\otimes |N,V_{12}\rangle=| N+1, V_{12}\rangle.
\end{equation}
Then, based on this result and Eq.\eqref{NNNNNN}, we can calculate some of the other Clebsh-Gordan coefficients. We have
\begin{eqnarray}\label{N1212}
&&\langle N, V_{12};1, V_{12}|N+1, V_{12}\rangle \langle N+1, \mathbf{0}|N, \mathbf{0};1, \mathbf{0}\rangle=\langle N+1, \mathbf{0}|N, \mathbf{0};1, \mathbf{0}\rangle\\\nonumber
& =&\sqrt{\frac{\dim(\pi_{N+1})}{(D+1)\cdot\dim(\pi_{N}) }}
\int_{SO(D+1)} d\bm{x}\overline{\Xi^{N+1,V_{12}}(\bm{x})} {\Xi^{N,V_{12}}(\bm{x}) }{\Xi^{1,V_{12}}(\bm{x})},
\end{eqnarray}
and
\begin{eqnarray}\label{N1221}
&&\langle N+1, V_{12};1, V_{21}|N, V_{12}\rangle \langle N, \mathbf{0}|N+1, \mathbf{0};1, \mathbf{0}\rangle\\\nonumber
& =&\sqrt{\frac{\dim(\pi_{N})}{(D+1)\cdot\dim(\pi_{N+1}) }}
\int_{SO(D+1)} d\bm{x}\overline{\Xi^{N,V_{12}}(\bm{x})} {\Xi^{N+1,V_{12}}(\bm{x}) }{\Xi^{1,V_{21}}(\bm{x})}.
\end{eqnarray}
Note that $\Xi^{1,V_{21}}(\bm{x})=\overline{\Xi^{1,V_{12}}(\bm{x})}$, then Eqs.\eqref{N1212} and \eqref{N1221} give that
\begin{equation}
\langle N, V_{12}|N+1, V_{12};1, V_{21}\rangle=\frac{\dim(\pi_{N})}{\dim(\pi_{N+1}) }\frac{\langle N+1, \mathbf{0}|N, \mathbf{0};1, \mathbf{0}\rangle}{\langle N+1, \mathbf{0};1, \mathbf{0}|N, \mathbf{0}\rangle}.
\end{equation}
Let us take
\begin{eqnarray}\label{CG0}
|1, V_{21};N,V_{12}\rangle :=|1, V_{21}\rangle\otimes |N,V_{12}\rangle &=&\alpha_1(N)| N-1, V_{12}\rangle +\alpha_2(N)|\text{other}\rangle,
\end{eqnarray}
where $|\text{other}\rangle$ represent a normalized state which are orthogonal with $ | N'', V_{12}\rangle, \forall N''\in\mathbb{N}$. Then we have
\begin{eqnarray}
|\alpha_1(N)|^2&=&\left|\frac{\dim(\pi_{N-1})}{\dim(\pi_{N}) }\frac{\langle N, \mathbf{0}|N-1, \mathbf{0};1, \mathbf{0}\rangle}{\langle N, \mathbf{0};1, \mathbf{0}|N-1, \mathbf{0}\rangle}\right|^2\\\nonumber
&=&
\frac{N(2N+D-3)}{(D+N-2)(2N+D-1)}.
\end{eqnarray}
and $|\alpha_2(N)|^2=1-|\alpha_1(N)|^2$.
Similar  discussion for other states gives
\begin{eqnarray}\label{CG1}
|1, V_{\imath\jmath}\rangle\otimes |N,V_{12}\rangle&=&|\text{other}'\rangle,\ \  (\imath,\jmath)=(3,4),(4,3),...\\\nonumber
{|1, \delta_{D+1}\rangle}\otimes |N,V_{12}\rangle&=&|\text{other}''\rangle,
\end{eqnarray}
where  $|\text{other}'\rangle$ and $|\text{other}''\rangle$ represent some normalized states which are orthogonal with $ | N'', V_{12}\rangle, \forall N''\in\mathbb{N}$.

Without loss of generality, we consider the holonomy operator $\hat{h}$ acting on the matrix element function $\Xi^{N}_{u^{-1},\tilde{u}}(h):=\langle N,V_{12}|u^{-1}h\tilde{u}|N,V_{12}\rangle$. 
 Then, the action of the operator corresponding to the holonomy component $\langle 1, V_{\imath\jmath}|u^{-1}h\tilde{u}|1, V_{\imath'\jmath'}\rangle$ can be given as
\begin{eqnarray}
\widehat{\langle 1, V_{\imath\jmath}|u^{-1}{h}\tilde{u}|1, V_{\imath'\jmath'}\rangle} \circ\Xi^{N}_{u^{-1},\tilde{u}}(h)&:=&\langle 1, V_{\imath\jmath}|u^{-1}h\tilde{u}|1, V_{\imath'\jmath'}\rangle \cdot \Xi^{N}_{u^{-1},\tilde{u}}(h)\\\nonumber
&=&\langle 1, V_{\imath\jmath}|u^{-1}h\tilde{u}|1, V_{\imath'\jmath'}\rangle \cdot \langle N,V_{12}|u^{-1}h\tilde{u}|N,V_{12}\rangle.
\end{eqnarray}
Based on this action, the matrix elements of the operator $\widehat{\langle 1,V_{\imath\jmath}|u^{-1}h\tilde{u}|1, V_{\imath'\jmath'}\rangle}$ in the basis spanned by the states $\{\left|N,u^{-1},\tilde{u}\right\rangle:=\langle N,V_{12}|u^{-1}h\tilde{u}|N,V_{12}\rangle|\}$ can be given by
\begin{eqnarray}\label{holoop}
&& \left\langle N',u^{-1},\tilde{u}\left|\widehat{\langle 1, V_{\imath\jmath}|u^{-1}h\tilde{u}|1, V_{\imath'\jmath'}\rangle}\right|N,u^{-1},\tilde{u}\right\rangle
  \\\nonumber&:=&\int_{SO(D+1)}dh \overline{\langle N',V_{12}|u^{-1}h\tilde{u}|N',V_{12}\rangle} \cdot \langle 1, V_{\imath\jmath}|u^{-1}h\tilde{u}|1,V_{\imath'\jmath'}\rangle \cdot \langle N,V_{12}|u^{-1}h\tilde{u}|N,V_{12}\rangle \\\nonumber
   &=&\frac{1}{\dim(\pi_{N'})}\langle1, V_{\imath\jmath};N,V_{12}|N',V_{12}\rangle\cdot \langle N',V_{12}|1, V_{\imath'\jmath'};N,V_{12}\rangle\\\nonumber
  &{=}& \frac{1}{\dim(\pi_{N'})}(\delta_\imath^{1}\delta_\jmath^{2}\delta_{\imath'}^{1}\delta_{\jmath'}^{2}\delta_{N',N+1}  +\delta_{N',N-1}\delta_\imath^{2}\delta_\jmath^{1}\delta_{\imath'}^{2}\delta_{\jmath'}^{1}|\alpha_1(N)|^2 ),
\end{eqnarray}
where we used Eqs.\eqref{CG00}, \eqref{CG0}, \eqref{CG1} and $\dim(\pi_N)=\frac{(D+N-2)!(2N+D-1)}{(D-1)!N!}$.

\bibliographystyle{unsrt}

\bibliography{ref}

\end{document}